\documentclass[a4paper,11pt]{article}

\usepackage[a4paper,top=1in,bottom=1in,left=1in,right=1in,marginparwidth=2cm]{geometry}

\usepackage[toc,page]{appendix}
\usepackage{cite} 
\usepackage{url} 
\usepackage[utf8]{inputenc} 
\usepackage{booktabs} 
\usepackage{graphicx}
\usepackage[scr=boondox]{mathalfa}
\usepackage{amsmath}
\usepackage{amsfonts}
\usepackage{amssymb}
\usepackage[table]{xcolor}
\usepackage{graphicx}
\usepackage{float}
\usepackage{caption}
\usepackage{lipsum}
\usepackage{titlesec}
\usepackage{multirow}
\usepackage{jinstpub}
\usepackage{tablefootnote}
\usepackage{lineno}
\usepackage{tipa}
\graphicspath{{plots/}}
\usepackage[separate-uncertainty = true,multi-part-units=single]{siunitx}
\usepackage{listings}
\newcommand{\thicc}{\mathscr{t}}
\newcommand{\pitch}{p}
\newcommand{\Bunit}{(\kilo\volt\per MeV) (\gram\per \milli\liter)}
\newcommand{\altgamma}{\textit{\textgamma}}
\newcommand{\ppid}{$p$-like PID\ }
\newcommand{\upid}{$\mu$-like PID\ }

\title{Angular dependent measurement of electron-ion recombination in liquid argon for ionization calorimetry in the ICARUS liquid argon time projection chamber}
\author[28]{P. Abratenko}
\author[4]{N. Abrego-Martinez}
\author[7]{A. Aduszkiewicz}
\author[23]{F. Akbar}
\author[27]{L. Aliaga Soplin}
\author[17]{M. Artero Pons}
\author[27]{J. Asaadi}
\author[6]{W. F. Badgett}
\author[17]{B. Baibussinov}
\author[5]{B. Behera}
% ...
\affiliation[4]{Centro de Investigacion y de Estudios Avanzados del IPN (Cinvestav), Mexico City}
\affiliation[5]{Colorado State University, Fort Collins, CO 80523, USA}
\affiliation[6]{Fermi National Accelerator Laboratory, Batavia, IL 60510, USA}
\affiliation[7]{University of Houston, Houston, TX 77204, USA}
% ...
\affiliation[17]{INFN Sezione di Padova and University, Padova, Italy}
% ...
\affiliation[23]{University of Rochester, Rochester, NY 14627, USA}
% ...
\affiliation[27]{University of Texas at Arlington, Arlington, TX 76019, USA}
\affiliation[28]{Tufts University, Medford, MA 02155, USA}
% ...
\emailAdd{gputnam@fnal.gov}
\date{\today}
\abstract{This paper reports on a measurement of electron-ion recombination in liquid argon in the ICARUS liquid argon time projection chamber (LArTPC). A clear dependence of recombination on the angle of the ionizing particle track relative to the drift electric field is observed. An ellipsoid modified box (EMB) model of recombination describes the data across all measured angles. These measurements are used for the calorimetric energy scale calibration of the ICARUS TPC, which is also presented. The impact of the EMB model is studied on calorimetric particle identification, as well as muon and proton energy measurements. Accounting for the angular dependence in EMB recombination improves the accuracy and precision of these measurements.}

\begin{document}
\maketitle
\flushbottom

\section{Introduction}
\label{sec:Introduction}
Liquid argon time projection chamber (LArTPC) detectors collect electrons ionized from argon by charged particles
for use in tracking and calorimetry \cite{Carlo}. Ionization electrons are drifted from the point of production by a large electric field to a set of readout wire planes which measure the charge. Critically, not all of the ionization electrons escape the particle track to be detected. Depending on the charge density, a significant fraction recombines with argon ions at the point of creation  \cite{ImelRecomb, ImelRecombStat, ICARUSrecomb, NEUTRecomb}. The rate of recombination has a non-linear dependence on the energy per length, or $dE/dx$, deposited by charged particles. In addition, because the drift electric field points in a specific direction, the recombination process may depend on the angle of the ionizing track to the electric field \cite{Columnar, Ellipsoid}. Measuring the rate of recombination across relevant variables is necessary for LArTPC detectors to precisely leverage calorimetry for particle identification and energy reconstruction. Any angular dependence in recombination is also of interest for argon-based dark matter detectors, where it could be leveraged to identify weakly interacting massive particle dark matter below the neutrino floor \cite{ArDM}. 

This paper reports on a measurement of electron-ion recombination and its dependence on the angle of the ionizing particle direction to the drift electric field with the ICARUS LArTPC neutrino detector \cite{ICARUSOG, ICARUSOverhaul}. The measurement is applied in the absolute energy scale calibration of the ICARUS TPC.
The measurement is performed by fitting for the electronics gain and recombination parameters in a single, self-consistent fit. This fit includes minimum-ionizing depositions from cosmic-ray muons (which are included in the fit to anchor the gain) and highly-ionizing depositions from protons produced in neutrino interactions at ICARUS from the Neutrinos at the Main Injector (NuMI) beam \cite{NuMI} (which provide information on the non-linearity of recombination). 

\begin{figure}[]
    \centering
    \includegraphics[width=\textwidth]{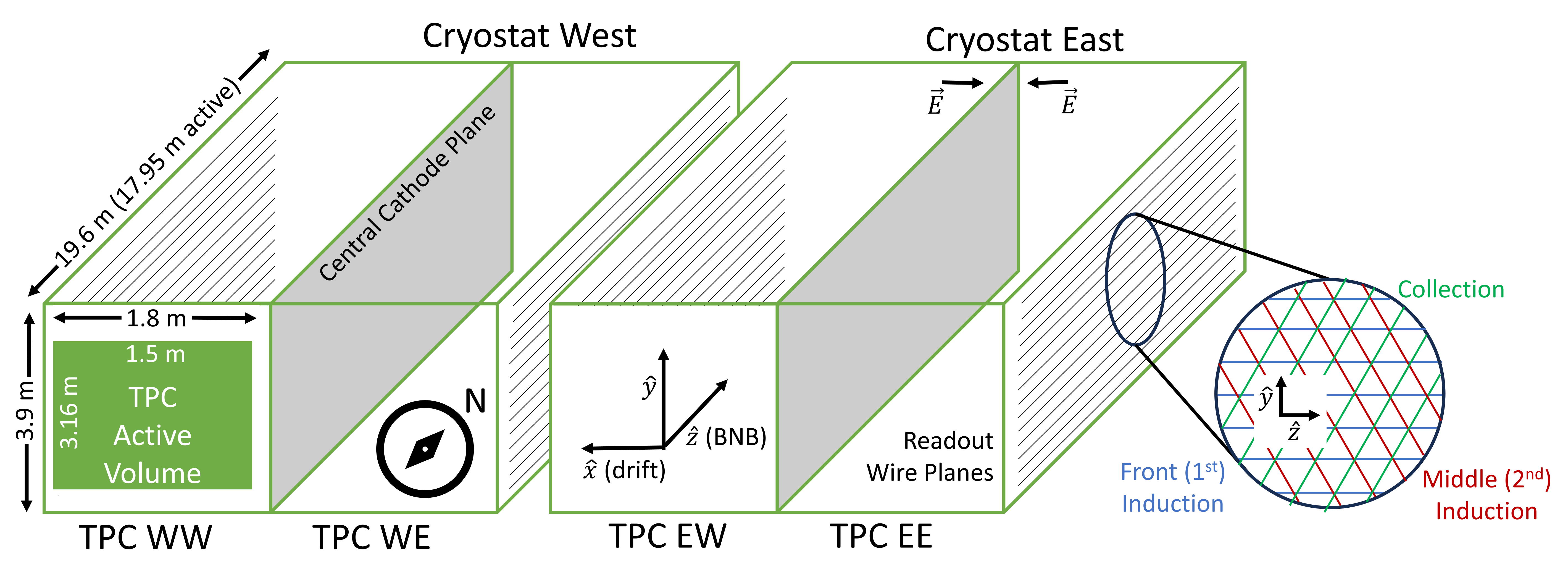}
    \caption{Diagram of the layout and enumeration of the ICARUS TPCs. Not to scale. The wire plane orientations are mirrored in opposite TPCs. The East and West cryostats have the same layout.}
    \label{fig:TPCDiagram}
\end{figure}

ICARUS is currently taking data as part of the Short-Baseline Neutrino Program \cite{SBNProposal, SBNProgram}. The ICARUS detector consists of two cryostats each with two TPCs separated by a common central cathode plane. All four TPCs are operated at a drift voltage of about \SI{500}{\volt\per\centi\meter}. The TPCs all have three planes of charge sensing wires: an unshielded front induction plane, a middle induction plane, and a collection plane. The wires on the front induction plane are oriented along the horizontal (beam) direction, and the wires on the middle induction and collection plane are oriented at $\pm 60^\circ$ to the horizontal direction, depending on the TPC. The results of this paper all use exclusively charge measurements on the collection plane. A diagram of the layout and enumeration of the four ICARUS TPCs is shown in figure \ref{fig:TPCDiagram}. The data used for this measurement is taken from ICARUS Run 1, which spanned from June 9 to July 9, 2022.

Each ICARUS TPC is instrumented to collect and digitize ionization charge with a linear gain and minimal noise \cite{ICARUSOverhaul, ICARUSElectronics}. Digitized waveforms are run through signal processing algorithms to further reduce noise and produce a Gaussian shape for ionization charge. A hit-finding algorithm identifies pulses on waveforms above a threshold and extracts the charge in the hit \cite{ICARUSResults}. Charge measurements in ICARUS are equalized to remove non-uniformities in space and time. ICARUS simulation also includes a robust model of the TPC noise and signal shapes to accurately describe charge reconstruction performance \cite{ICARUSSignalNoise}. 

This paper is organized as follows. Section \ref{sec:TrackEnergyScale} discusses the expected energy loss for proton and muon tracks used in the calibration. Section \ref{sec:Recombination} discusses the models used to describe recombination and how to include an angular dependence. Section \ref{sec:TrackSelection} develops how stopping muon and proton tracks are selected. Section \ref{sec:EnergyFit} shows the fit results for the TPC gain and the angular dependent recombination measurement. The fit is applied as the energy scale calibration for ionization calorimetry in ICARUS. Applying this calibration, section \ref{sec:ParticleID} shows results for particle identification and calorimetric energy measurements and compares data to Monte Carlo simulation. Finally, section \ref{sec:Conclusion} concludes the paper.


\section{Particle Track Energy Scale}
\label{sec:TrackEnergyScale}
The energy scale calibration is determined by fitting the observed ionization charge per length ($dQ/dx$) to the expected energy loss per length ($dE/dx$) along the particle trajectory. Muon and proton energy loss in the kinematic regime relevant to ICARUS is primarily by ionization. The mean energy loss from ionization is given by Bethe-Bloch theory \cite{Bethe}, with \cite{PDG}
\begin{equation}
    \begin{split}
    \overline{\frac{dE}{dx}} &= \zeta T_\text{max}\left[\text{ln}\frac{2m_e c^2\beta^2 \altgamma^2 T_\text{max}}{I_0^2} - 2\beta^2 - \delta(\beta\altgamma)\right] \\
        T_\text{max} &= \frac{2m_e c^2 \beta^2\altgamma^2}{1 + 2\altgamma m_e/M + (m_e/M)^2}\\
        \zeta &= \rho\frac{K}{2}\frac{Z}{A}\frac{1}{T_\text{max}\beta^2}\,.
        \label{eq:dEdxmean}
    \end{split}
\end{equation}
The relevant parameters are the particle (muon or proton) mass $M$, the electron mass $m_e$, the particle velocity $\beta$, the Lorentz factor $\altgamma$, the mean excitation energy $I_0$, the argon charge number $Z$, the argon mass number $A$, the argon mass density $\rho$, and the  constant $K$, with units \si{MeV\times\centi\meter\squared \per\mol}. In these equations, $T_\text{max}$ is the maximum energy transfer to a single electron, $\zeta$, with units of inverse length, encodes the rate of scattering, and $\delta$ is the correction from the density effect \cite{DensityEffect}. We use the parameterization \cite{StoppingPower}
\begin{equation}
\delta(\beta\altgamma) = \begin{cases}
    0 &\text{log}_{10}\beta\altgamma < 0.2\\
    2\text{ln}\beta\altgamma - 5.2146 + 0.19559 \times (3 - \text{log}_{10}\beta\altgamma)^3 &0.2 \leq \text{log}_{10}\beta\altgamma \leq 3\\
    2\text{ln}\beta\altgamma - 5.2146 &  \text{log}_{10}\beta\altgamma > 3\, .
\end{cases}
\end{equation}

\begin{figure}[t]
    \centering
    \includegraphics[width=0.65\textwidth]{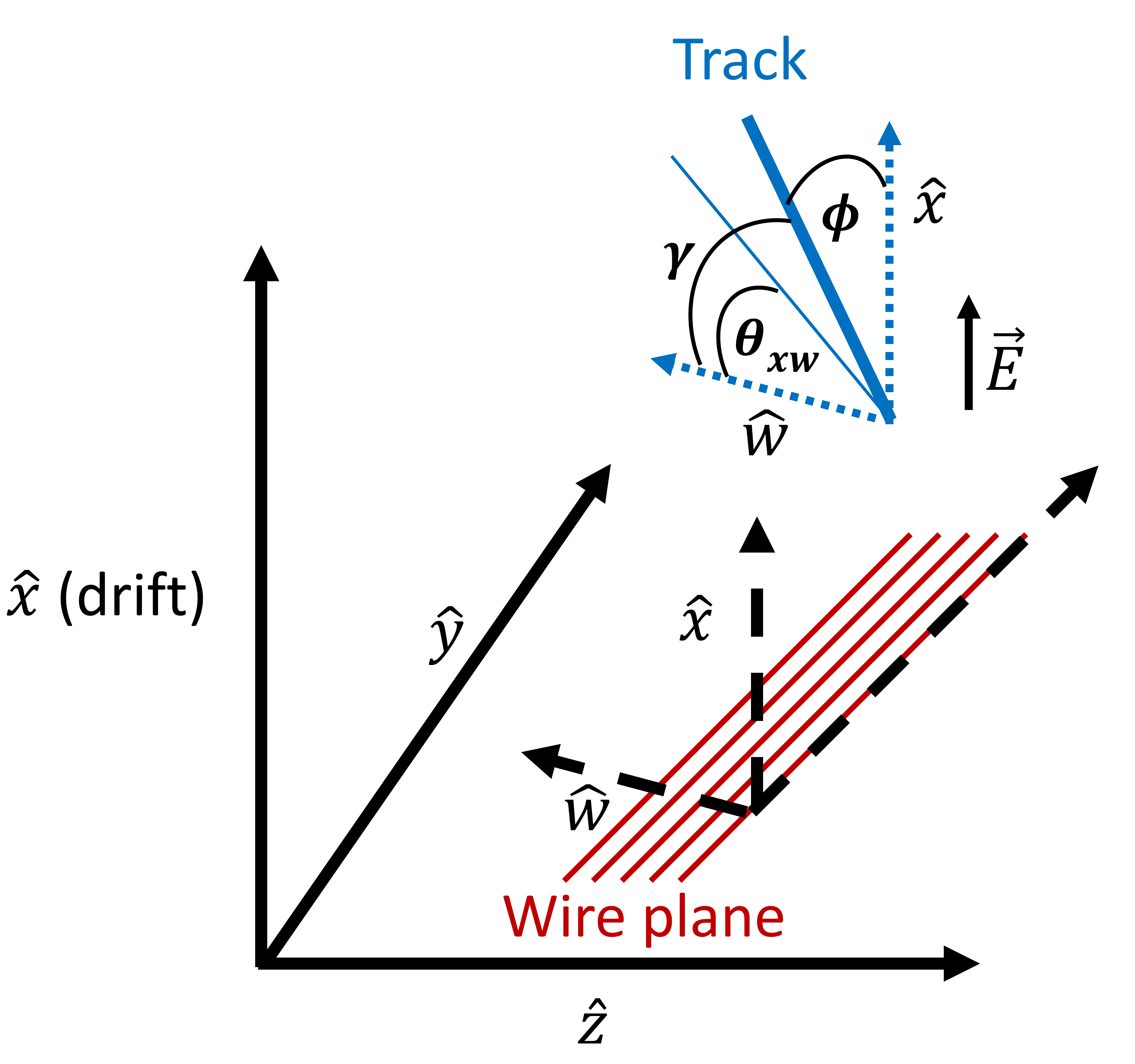}
    \caption{Diagram of relevant track angles. $\gamma$, the angle of the the track to the direction perpendicular to the wire direction, determines the track pitch. $\phi$, the angle of the track to the drift electric field, controls any angular dependence in electron-ion recombination. $\theta_{xw}$, the angle of the track between the drift electric field direction and the direction perpendicular to the wire direction, controls the track ionization signal shape.}
    \label{fig:anglediagram}
\end{figure}

However, due to the large fluctuations in particle energy loss as described by Landau-Vavilov theory \cite{Landau, Vavilov}, the mean energy loss is challenging to measure. Instead, we calibrate to the most-probable-value (MPV) of energy loss, which only depends on the peak of the distribution. In the Landau limit, which is applicable to energy depositions far from the particle track Bragg peak, this is given by \cite{PDG}
\begin{equation}
        \frac{dE}{dx}\bigg\rvert_\text{MPV}
        = \overline{\frac{dE}{dx}} + \zeta T_\text{max} \left( \text{log}[\zeta\thicc] + 0.2 + \beta^2\right),
        \label{eq:dEdxMPV}
\end{equation}
where $\thicc$ is the
length of the muon observed by the wire (the track ``thickness"). This effective length is a function of the track angle, as well as transverse diffusion. It is given by \cite{GrayDiff}
\begin{equation}
    \label{eq:thicc}
    \begin{split}
        \thicc(t_\text{drift}, \gamma) &= \frac{\pitch}{\text{cos}\gamma}\  \text{exp}\left(-\int \frac{dx}{\pitch} w[\sigma_T(t_\text{drift}), x]\  \text{log}\, w[\sigma_T(t_\text{drift}), x] \right)\\ 
        \sigma_T(t_\text{drift}) &= \sqrt{2 D_T t_\text{drift}}\\
        w(\sigma_T, x) &= \int\limits_{-\pitch/2}^{\pitch/2} \frac{dx'}{\sigma_T \sqrt{2\pi}} e^{-\frac{(x-x')^2}{2\sigma_T^2}},
    \end{split}
\end{equation}
where $\gamma$ is the angle of the track to the direction perpendicular to the wire orientation (see figure \ref{fig:anglediagram}), $\pitch$ is the wire pitch, and $D_T$ is the transverse diffusion coefficient. Previous LArTPC experiments have calibrated using the track pitch $\pitch/\text{cos}\gamma$ as the value of $\thicc$. This can cause a bias in the measured gain, as well as other measurements leveraging $dQ/dx$ \cite{GrayDiff, MichelleDiff}. 

The Landau approximation is quantified by the unitless quantity $\zeta \cdot \thicc$, and is typically taken to be valid when $\zeta \cdot \thicc < 0.01$ \cite{PDG}. When this limit is violated, the distribution of particle energy loss is not well modeled by a Landau distribution and equation \ref{eq:dEdxMPV} does not apply. The $\zeta$ parameter monotonically decreases with increasing momentum, so the approximation is valid at higher momentum, away from the Bragg peak.

In our calibration procedure, for the muon population we elect to only include depositions where the analytic Landau approximation is valid. This limits us to depositions near the mimumum-ionizing-particle (MIP) region, where recombination is close to linear. To provide information on recombination, we also include highly-ionizing-particle (HIP) depositions from protons near their Bragg peak. In this case, the distribution of energy loss is approximately a Landau-Vavilov distribution \cite{Vavilov}, although there is a perturbation to the shape from diffusion \cite{GrayDiff}. The MPV of the Landau-Vavilov distribution cannot be expressed analytically. We use the numerical computation provided by the ROOT VavilovAccurate routine\footnote{We have found in our simulation that in this region the MPV of the proton energy loss distribution is modeled well by the ROOT numerical computation, using the track pitch (not the track thickness) as the input to the Landau parameter ($\lambda_L$) and $\kappa$ value used in the VavilovAccurate routine.} \cite{ROOT}.

\begin{table}[t]
\centering
\begin{tabular}{ l | r } 
Parameter & Value \\
\hline
Energy loss coefficient $K$ \cite{PDG} & \SI{0.307075}{MeV\centi\meter\squared\per\mole}\\
Mean excitation energy $I_0$ \cite{Strait} & \SI{197 \pm 7}{eV}\\
Transverse diffusion constant $D_T$\tablefootnote{This value is taken from an analysis in ICARUS in preparation to be published. The preliminary value applied here will not change in the final result by an amount significant enough to impact the results of this paper.} & \SI{7.5 \pm 0.2}{\centi\meter\squared\per\second}\\
Argon ionization work function $W_\text{ion}$ \cite{Wion} & 23.6 $\pm$ \SI{0.3}{eV}\\
Argon density $\rho$ & \SI{1.393}{\gram\per\milli\liter}\\
Drift electric field $\mathcal{E}$ & 492.6 $\pm$ \SI{8.4}{\volt\per\centi\meter}\\
Wire pitch $\pitch$ \cite{ICARUSOG} & \SI{2.991}{\milli\meter}\\
\end{tabular}
\caption{Numerical values of parameters that determine the muon and proton charge scale. A reference is included for external measurements. Uncertainties are shown when their size is relevant.}
\label{tbl:values}
\end{table}

The numerical values we have used to compute the muon and proton energy scale are listed in table \ref{tbl:values}. The largest uncertainty in any parameter relevant to energy loss modeling comes from the mean excitation energy. We do not specifically include an uncertainty from this parameter in the energy scale fit, as other uncertainties in the fit dominate over the impact of the excitation energy on the mean energy loss. 

\section{Recombination Modeling}
\label{sec:Recombination}
Electron-ion recombination in liquid argon is driven by the collective absorption of ionized electrons by the cloud of argon ions along the particle track \cite{ImelRecomb, NEUTRecomb} (the geminate fraction is negligible \cite{RecombinationGerminate}). Recombination occurs after electrons thermalize and before they diffuse significantly \cite{RecombinationThermal}. It happens before the argon ions drift or diffuse significantly. Recombination thus takes place while the ionization electrons are dragged by the drift electric field over a stationary cloud of argon ions. The rate of recombination therefore may depend on the strength of the drift electric field and its orientation to the particle track, as well as the particle stopping power ($dE/dx$).

We consider two models to parameterize the recombination of ionizing particle tracks in liquid argon. The previous ICARUS measurement of recombination at Gran Sasso applied the Birks equation \cite{ICARUSrecomb}
\begin{equation}
\label{eq:birks}
        \frac{dQ}{dx} = \frac{1}{W_\text{ion}}\frac{A\frac{dE}{dx}}{1 + k\frac{dE}{dx}/\mathcal{E}\rho},
\end{equation}
where $W_\text{ion}$ is the argon ionization work function, $\mathcal{E}$ is the drift electric field, and $A$ and $k$ are fit parameters. The ArgoNeuT experiment measured recombination in terms of their proposed modified box model \cite{NEUTRecomb}
\begin{equation}
\label{eq:modbox}
    \begin{split}
        \frac{dQ}{dx} &= \frac{\text{log}\left(\alpha + \mathcal{B}\frac{dE}{dx}\right)}{\mathcal{B}W_\text{ion}}\\
        \mathcal{B} &= \frac{\beta}{\mathcal{E}\rho},
    \end{split}
\end{equation}
where $\alpha$ and $\beta$ are fit parameters. This equation is a modification of the Thomas-Imel box model \cite{ImelRecomb}.

In both of these models one fit parameter ($A, \alpha$) is uncoupled to the electric field. At the ICARUS drift field, these parameters control the amount of recombination for minimum-ionizing depositions. Both models also include one parameter ($k, \beta$) which is coupled to the electric field and determines the non-linearity of recombination with respect to $dE/dx$.

Neither of these models explicitly include a dependence on the track angle to the drift field, which we refer to as $\phi$ (see figure \ref{fig:anglediagram}).  Different forms of the angular dependence derive from different assumed shapes for the particle track ionization cloud. Two such examples are columnar \cite{Columnar} and ellipsoid \cite{Ellipsoid} shapes. These models give an angular dependence to the drift electric field in the Birks equation. The same angular dependence in the drift electric field can also be applied in the modified box model.

For this measurement, we elect to include an angular dependence with a general form by promoting the parameters coupled to the drift field, $k$ for Birks recombination and $\beta$ for modified box recombination, to functions of $\phi$: $k(\phi)$ and $\beta(\phi)$. These are compared to the columnar and ellipsoid model predictions. Neither phenomenological model assigns any angular dependence to the parameters uncoupled to the electric field ($A, \alpha$). We do not include any angular dependence in these parameters in our measurement. We find that our results are well described by including an angular dependence just on the parameters coupled to the electric field ($k, \beta$).

Our recombination modeling also assumes that there is no dependence of recombination on the particle type (muon versus proton, e.g.). It is possible that at the same energy loss ($dE/dx$), muons and protons would deposit different amounts of ionization ($dQ/dx$) because the two particles produce different energy spectra of ionization electrons. Our results are well described by neglecting any particle type dependence, although they do not include any data where the expected muon and proton energy loss overlaps. The search for such an effect merits further study.

In the fit, we also include the electronics gain ($\mathcal{G}$) as a free parameter. The gain enters the fit equations as $\frac{dQ}{dx}_\text{ADC} = \frac{1}{\mathcal{G}}\frac{dQ}{dx}$. The gain for the ICARUS TPC readout electronics has been previously measured \cite{ICARUSElectronics}. The gain in this fit should be understood as an effective parameter which encodes any perturbations induced by signal processing and charge corrections. In addition, by including the gain directly in the fit, we are able to include the uncertainty that the unknown effective gain induces on the measurement of recombination.  

\section{Track Selection}
\label{sec:TrackSelection}
Particle tracks are identified in ICARUS with the Pandora reconstruction framework \cite{Pandora, DUNEPandora}, run on hits produced by the ICARUS signal processing chain \cite{ICARUSResults}. Separate track selections identify cosmic-ray muon and neutrino-induced protons useful for the energy scale calibration. These are detailed in sections \ref{sec:muonselection} and \ref{sec:protonselection}, respectively. The charge measured from tracks is equalized across space, time, and track orientation, as described in section \ref{sec:chargecorr}. Distributions of $dQ/dx$ are then constructed from hits along tracks, as discussed in section \ref{sec:chargedist}. Table \ref{tbl:selection} outlines the selection for proton and muon hits.

\begin{table}[t]
\centering
\rowcolors{2}{gray!25}{white}
\begin{tabular}{ l | p{4cm} | p{5cm}} 
Selection Step & Muons & Protons \\
\hline
Topological Selection & Cathode-crossing (see section \ref{sec:muonselection}) & From neutrino candidate (see section \ref{sec:protonselection})\\
Calorimetric Selection & Median $dQ/dx$ in last \SI{5}{cm} > 75 ke$^-/$cm & \upid $> 40$, \ppid $< 80$ (see figure \ref{fig:chi2up}) \\
Fiducial Volume X Inset (drift) & \multicolumn{2}{l}{\SI{10}{cm} from anode and \SI{15}{cm} from cathode} \\
Fiducial Volume Y Inset (vertical) & \multicolumn{2}{l}{\SI{20}{cm} from top and bottom. \SI{75}{cm} from top in TPC WW.} \\
Fiducial Volume Z Inset (BNB) & \multicolumn{2}{l}{\SI{50}{cm} from front and back}\\
Hit Residual Range & 33 bins, 80 to \SI{300}{cm} & 16 bins, 2 to \SI{25}{cm}\\
Hit Pitch & 0.3 to \SI{0.4}{cm} & 0.3 to \SI{1}{cm}\\
Hit $\phi$ & 70$^\circ$ to 85$^\circ$ & 6 bins, 30$^\circ$ to 85$^\circ$\\
Hit $\theta_{xw}$ & 5$^\circ$ to 20$^\circ$ & 5$^\circ$ to 70$^\circ$\\
Hit Drift Time & 5 bins, 500 to \SI{900}{\micro\second} & No cut\\
\end{tabular}
\caption{Overview of selection to identify hits from muon and proton tracks. The topological section is specified in sections \ref{sec:muonselection} and \ref{sec:protonselection} for muon and protons, respectively. The extra fiducial cut in TPC WW removes a problematic detector region. When hits are split into groups by the quantity, the number of bins is specified.}
\label{tbl:selection}
\end{table}

\subsection{Cosmic-Ray Muon Selection}
\label{sec:muonselection}

Cosmic-ray muon tracks are required to cross the central cathode plane in either cryostat. The identification of a track in both TPCs on either side of the cathode allows the arrival time of the muon to be determined. This enables the track charge to be corrected for attenuation due to argon impurities. These ``cathode-crossing tracks" are identified by Pandora. The muon is required to stop within a fiducial volume (see table \ref{tbl:selection}), and is required to have a median $dQ/dx$ in the last \SI{5}{\centi\meter} along the track greater than $\sim$ 75 ke$^-/$cm
\footnote{The cut value is expressed directly in terms of analog-digital counts (ADC) per centimeter. The cut is \SI{1000}{ADC\per\centi\meter}. The written value is converted to a number of electrons ($e^-$) applying an approximate gain of 75 $e^-/$ADC.}. 
This cut increases the stopping track purity by requiring the measured charge near the endpoint to be consistent with the expectation from the muon Bragg peak. The region of the track where the cut is applied is not included in the calibration, and so the cut should not bias the measurement. This selection identifies 93 thousand muon candidates. A Monte Carlo simulation study (with cosmic-ray muon generation modeled by CORSIKA \cite{Corsika}) of this selection yields a 93\% purity of tracks stopping within \SI{5}{\centi\meter} of the reconstructed endpoint. The remaining 7\% consist almost entirely of cosmic-ray muons where the endpoint is mis-reconstructed, either due to track splitting or combining with a Michel electron. The dependence of energy loss on the muon momentum is small for the range of muon momenta used in this measurement, so this impurity has a small impact.

\subsection{Neutrino-Induced Proton Selection}
\label{sec:protonselection}

Neutrino-induced protons are selected by applying topological and loose calorimetric cuts. The tracks are required to originate from a neutrino candidate with a fiducial interaction vertex. The neutrino candidate must have at least two tracks with a length of at least \SI{25}{\centi\meter} each. The cosine of the angle of the longest track in the candidate to the vertical direction must be greater than -0.7. These cuts select for event topologies with multiple tracks and remove events with downward going tracks. They remove the large majority of cosmic-ray muon background events.

\begin{figure}[t]
    \centering
    \includegraphics[width=\textwidth]{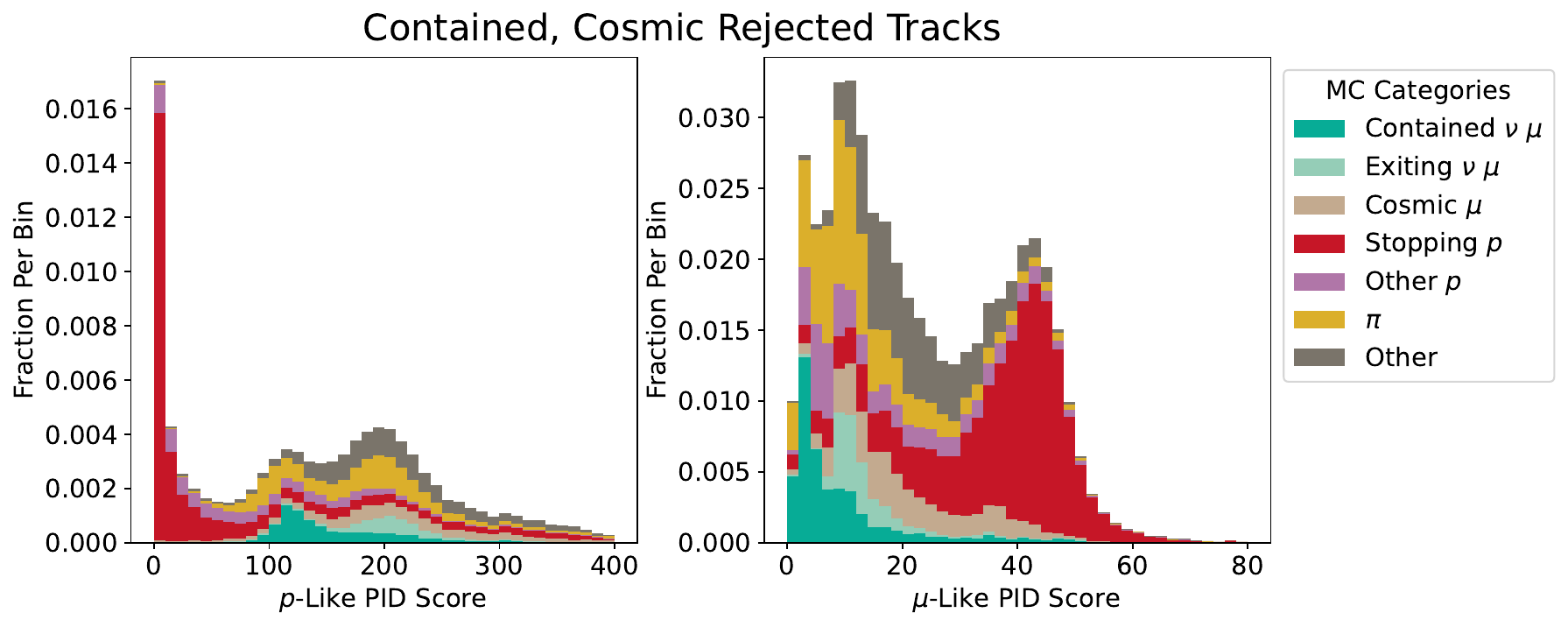}
    \caption{Distribution of proton-like (left) and muon-like (right) particle identification (PID) variables in ICARUS NuMI neutrino + CORSIKA cosmic-ray Monte Carlo simulation. Distributions are shown after applying the topological cuts in section \ref{sec:protonselection}. The particle ID variables are computed by comparing the profile of $dE/dx$ along a track to the theoretical expectation for the proton and muon hypotheses.}
    \label{fig:chi2up}
\end{figure}

After the topological cuts, the proton track candidates are selected by applying a calorimetric particle identification cut. This cut relies on PID variables which compare the reconstructed profile of $dE/dx$ along a track to the theoretical expectation for muon ($\mu$-like) and proton ($p$-like) hypotheses \cite{ArgoNeuTPID}. Calorimetric cuts are necessary to select proton tracks from muons and pions, and have been used in prior recombination measurements leveraging protons \cite{NEUTRecomb, CalibrationuB}. These variables are computed using a basic energy scale calibration which assumes the ArgoNeuT recombination measurement \cite{NEUTRecomb} and fits for the TPC electronics gain using the cosmic muon track sample. Distributions of the particle identification variables in ICARUS Monte Carlo simulation are shown in figure \ref{fig:chi2up}. Candidate proton tracks must be contained in a fiducial volume and have \ppid $< 80$ and \upid $> 40$. This selection identifies 4.4 thousand proton candidates. The sample has a 97.5\% purity of true protons in a Monte Carlo simulation study. We have performed the recombination measurement with variations on the calorimetric cuts and have found that our result is not sensitive to the particular choice of the cut values.

\subsection{Charge Scale Corrections}
\label{sec:chargecorr}

Charge signals measured from muon and proton tracks per-wire on the collection plane are equalized per-hit in space and time by applying the charge equalization procedure developed for ICARUS \cite{ICARUSSignalNoise}. This procedure equalizes the detector response by applying corrections obtained from charge depositions by thoroughgoing cosmic muons.

A small ($\sim$2.5\%) angular dependence is also observed in the charge reconstruction for particle tracks as derived by comparing charge reconstruction methods in ICARUS simulation and data. There are two methods for measuring charge in a hit in ICARUS reconstruction. The ``Integral" method fits a Gaussian shape to the hit pulse and defines the charge as the area of the fit. The ``SummedADC" method sums the ADC values over the range of the hit. The hit range is defined as the region between two local minima below the (baseline subtracted) zero-point of the waveform on either side of a hit pulse that goes above a set threshold. Monte Carlo simulations in ICARUS indicate that the SummedADC method has a worse charge resolution but no angular dependence, whereas the Integral method has a better resolution and a moderate angular dependence. The TPC signal shapes in ICARUS Monte Carlo simulation have been tuned to directly match the data signal shapes \cite{ICARUSSignalNoise}, so we are confident that the Monte Carlo simulation is able to precisely model the amount of angular dependence in the charge reconstruction.

\begin{figure}[]
    \centering
    \includegraphics[width=0.5\textwidth]{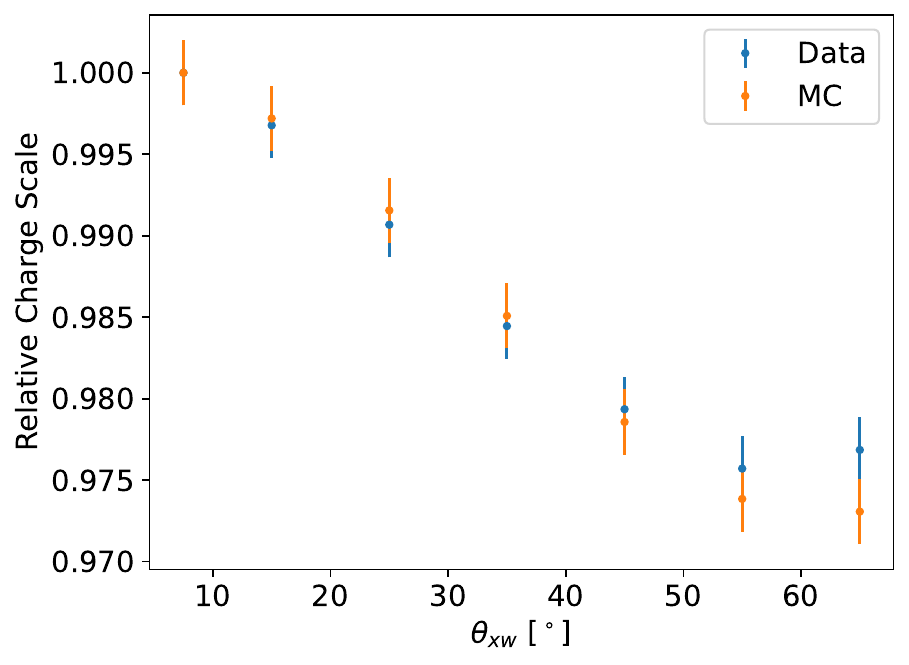}
    \caption{Relative scale of charge reconstruction as a function of the track angle $\theta_{xw}$ (see figure \ref{fig:anglediagram}), determined in ICARUS data and Monte Carlo simulation.  A systematic uncertainty of 0.2\% on the correction is assigned to cover the difference between data and Monte Carlo simulation.}
    \label{fig:dataangcorr}
\end{figure}

We use the Integral charge method for calibrating the ICARUS TPC energy scale, but use the SummedADC method to diagnose the angular dependence. The ratio of the Integral to SummedADC charge reconstruction is taken as a charge scale correction factor as a function of the track angle $\theta_{xw}$\footnote{This angle determines the shape of the ionization charge signal from a track, and therefore controls any orientation dependence in the charge reconstruction.} (see figure \ref{fig:anglediagram}). Figure \ref{fig:dataangcorr} plots the correction factor in Monte Carlo simulation and data. A systematic uncertainty of 0.2\% on the correction is assigned to cover the differences.

Selected protons are required to have $\theta_{xw} < 70^\circ$ so that this correction is applicable. For muons, instead of making a correction, we restrict the $\theta_{xw}$ range between $5^\circ < \theta_{xw} < 20^\circ$, over which the angular dependence is not significant ($<0.2\%$).

\subsection{$dQ/dx$ Measurements}
\label{sec:chargedist}

After selection and charge scale corrections, hits from protons and muons are divided into groups. The MPV $dQ/dx$ is extracted in each group. Both muon and proton hits are required to be contained in a fiducial volume which removes regions where drift electric field distortions are the largest. Muon hits are required to have a reconstructed pitch less than \SI{4}{\milli\meter} and a track angle $\phi > 70^\circ$. They are grouped by TPC, drift time, and residual range (the length of the muon after that hit). The subdivision by residual range and drift time selects for a single expected energy loss MPV in each distribution. The subdivision by TPC checks for any variations in the gain that were not removed by the charge equalization procedure. Proton hits are required to have a pitch less than \SI{1}{\centi\meter} and are grouped by residual range and the track angle $\phi$. See table \ref{tbl:selection} for the overview of the cuts applied to muon and proton hits.

The MPV $dQ/dx$ is obtained in each group by fitting the distribution of $dQ/dx$ values to a Landau distribution convolved with a Gaussian distribution. The fit includes a statistical uncertainty on the MPV.
Due to the uncertainty in the charge scale corrections to remove angular dependence (section \ref{sec:chargecorr}), as well as spatial and temporal variations \cite{ICARUSSignalNoise}, we also assign a 1\% systematic uncertainty on each proton $dQ/dx$ MPV. The systematic uncertainty is added in quadrature with the uncertainty from the fit. The uncertainty in the muon $dQ/dx$ MPV is separately validated by verifying that the fit uncertainty is consistent with the residuals of the MPVs to a linear fit to the data (since recombination is linear for the minimum-ionizing muon depositions). Thus, no additional systematic uncertainty is added to the muon $dQ/dx$ MPVs.

\section{Energy Scale Fit}
\label{sec:EnergyFit}
\subsection{Procedure}

The end results of the track and hit selection described in section \ref{sec:TrackSelection} are $dQ/dx$ MPVs for muon and proton tracks as a function of mean pitch, residual range, and (for muons only) drift time. Each $dQ/dx$ MPV is matched to an expected $dE/dx$ MPV by applying the energy scale modeling from section \ref{sec:TrackEnergyScale}. The measured $dQ/dx$ MPVs are fit to the expected $dE/dx$ MPVs with either the Birks or modified box recombination model. The electronics gain is included as a free parameter in both fits. The $A$ parameter in Birks recombination is degenerate with the gain in the fit, so we use the $A$ value from the ICARUS Gran Sasso measurement \cite{ICARUSrecomb}. The parameters coupled to the electric field, $k(\phi)$ for Birks recombination and $\beta(\phi)$ for modified box recombination, are allowed to be different in each proton angular bin. The muon data bridges two of the proton angular bins, $70^\circ < \phi < 80^\circ$ and $80^\circ < \phi < 85^\circ$. We use the average $\beta$ and $k$ values from those two bins to calculate the expected amount of recombination. We have verified in a Monte Carlo simulation study that this method is able to reproduce the simulated recombination and electronics gain parameters.

The external values of parameters used in the fit are listed in table \ref{tbl:values}. There is a 1.7\% uncertainty in the electric field which translates directly into an uncertainty in $\beta$ and $k$ which is fully correlated across angular bins. The uncertainty in the electric field arises from distortions due to space charge and bending of the cathode observed in both cryostats in ICARUS. 
There is an additional localized drift field distortion in one TPC (the East TPC in the East cryostat, TPC EE) due to a failure in the field cage. We have confirmed that removing data from that TPC does not change the measurement by more than the uncertainty on the drift electric field. We therefore include data from that TPC in the measurement.



\subsection{Results}

\begin{figure}[tp]
    \centering
    \includegraphics[width=0.49\textwidth]{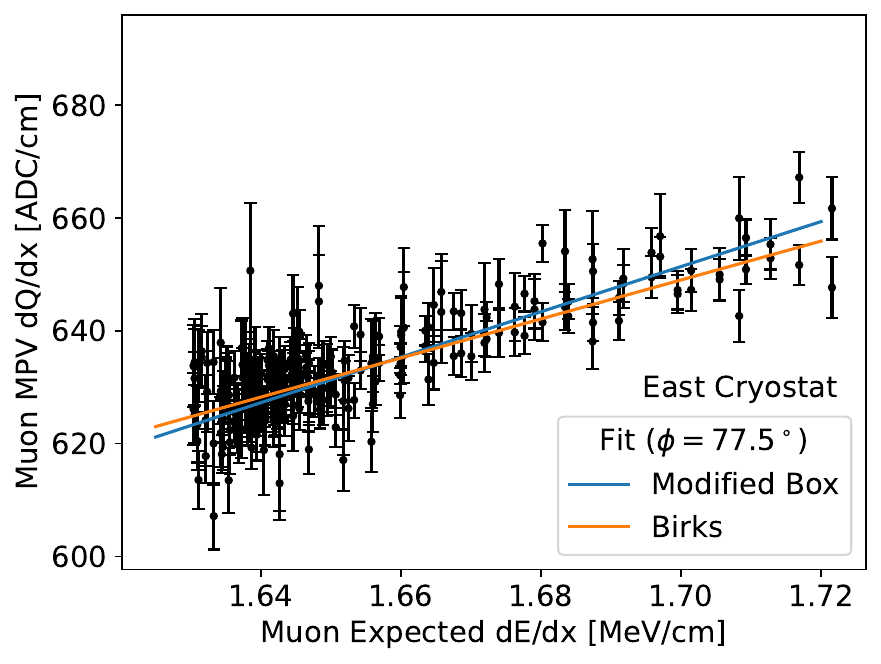}
    \includegraphics[width=0.49\textwidth]{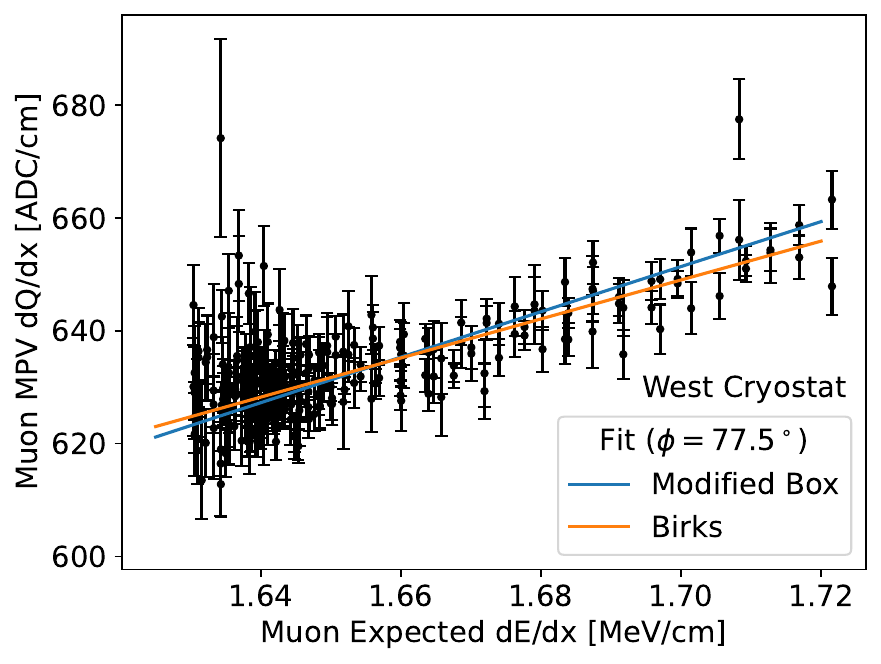}
    
    \caption{Fits of measured MPV $dQ/dx$ to expected MPV $dE/dx$ for muons in the East cryostat (left) and the West cryostat (right). The two lines compare the Birks and modified box fits.}
    \label{fig:muonfits}
\end{figure}

\begin{figure}[tp]
    \centering
    \includegraphics[width=0.4\textwidth]{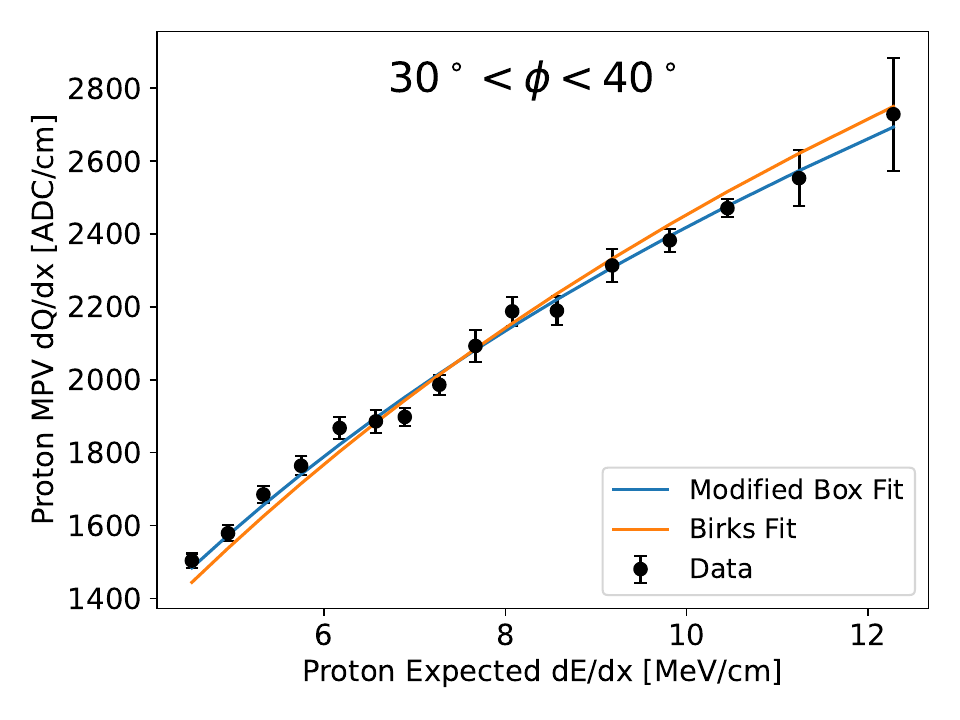}
    \includegraphics[width=0.4\textwidth]{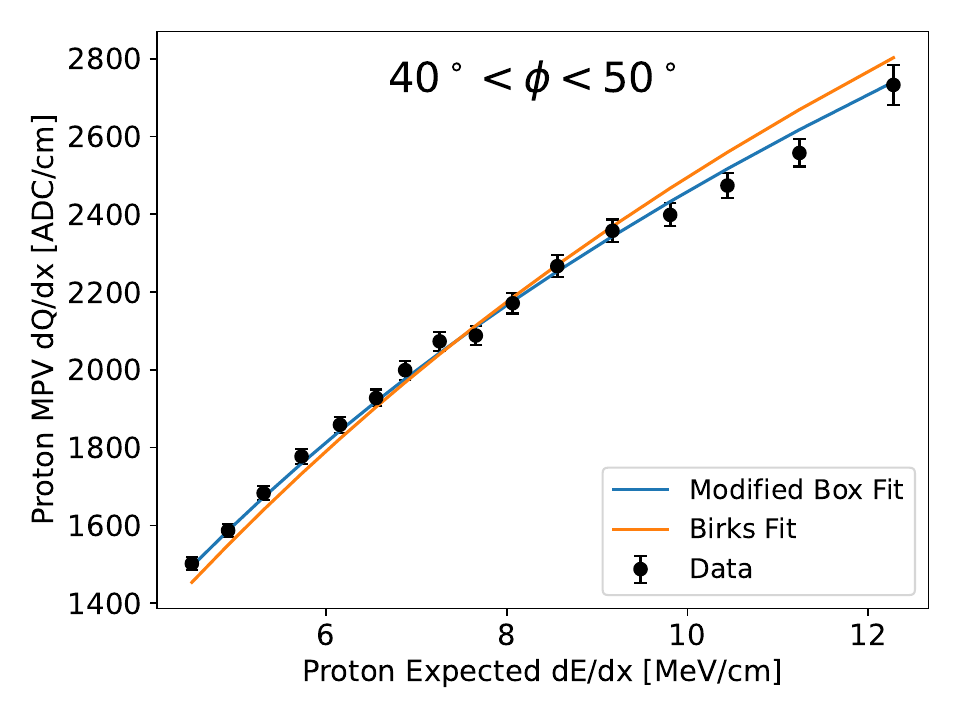}
    \includegraphics[width=0.4\textwidth]{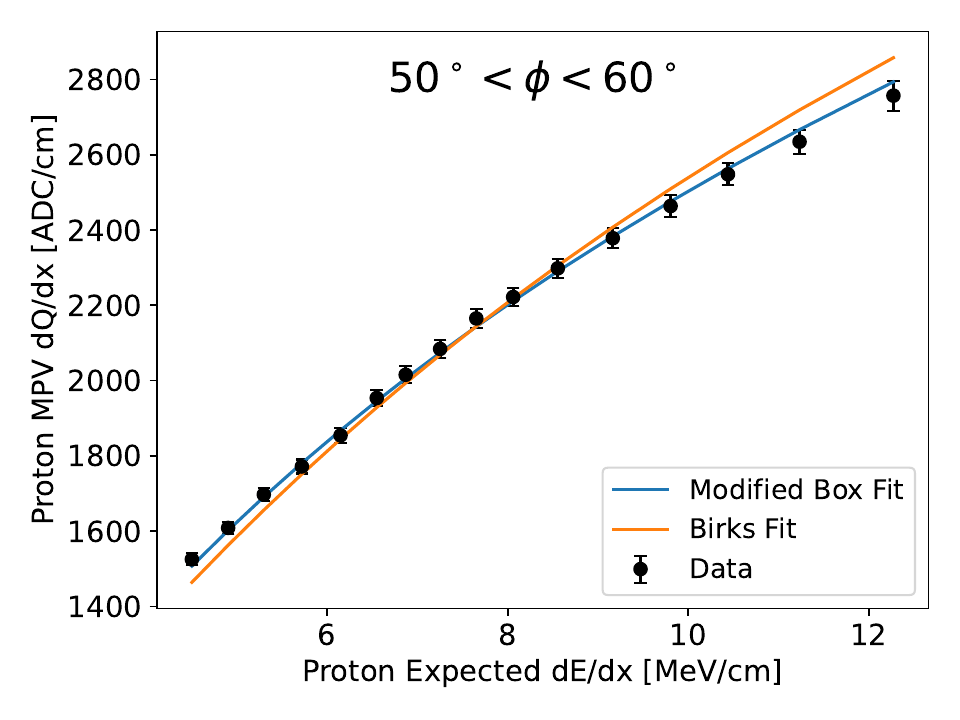}
    \includegraphics[width=0.4\textwidth]{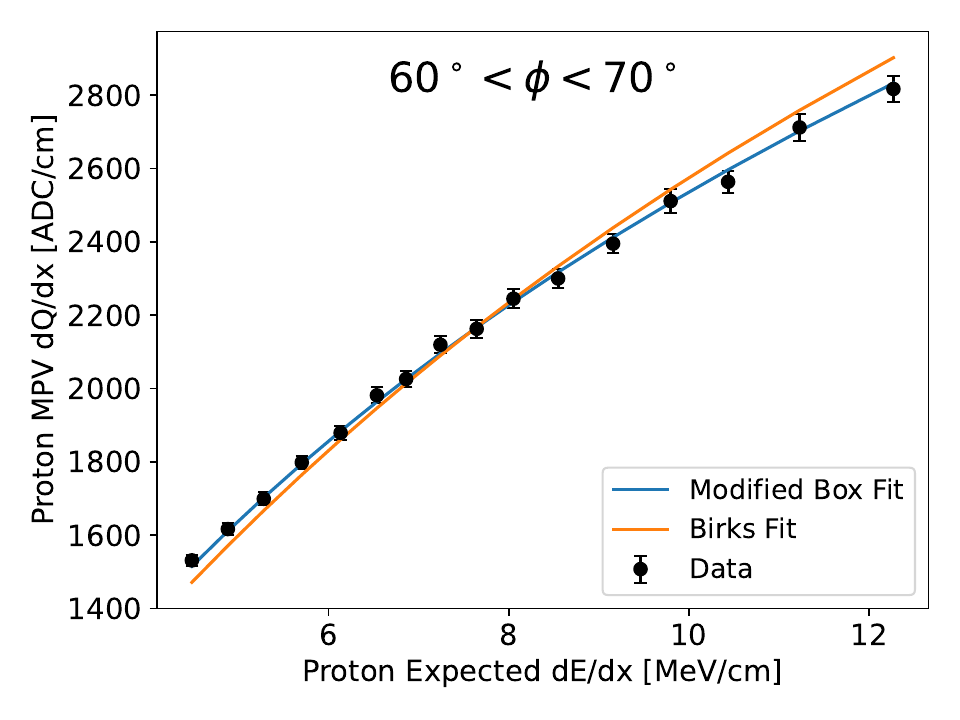}
    \includegraphics[width=0.4\textwidth]{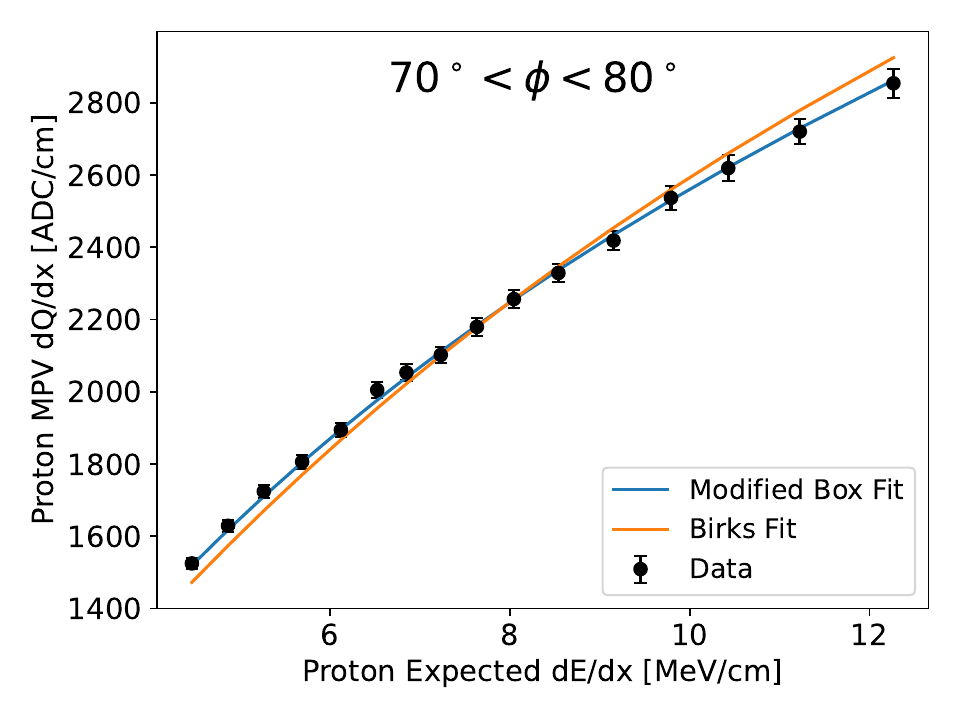}
    \includegraphics[width=0.4\textwidth]{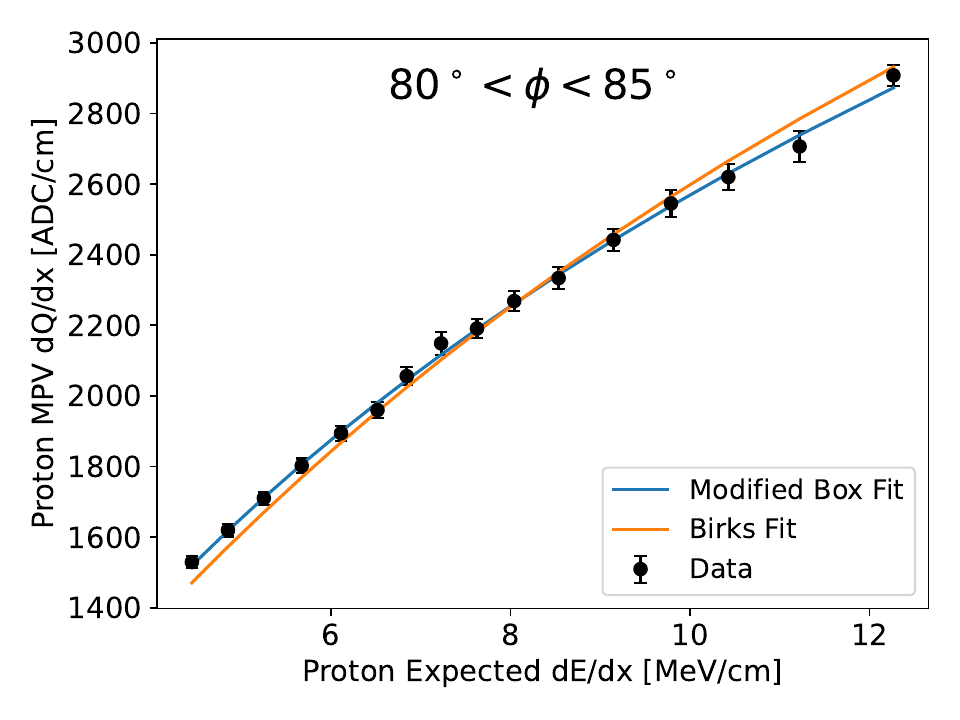}

    \caption{Fits of measured MPV $dQ/dx$ to expected MPV $dE/dx$ for protons. The two lines compare the Birks and modified box fits.}
    \label{fig:protonfit}
\end{figure}

\begin{figure}[t]
    \centering
    \includegraphics[width=0.49\textwidth]{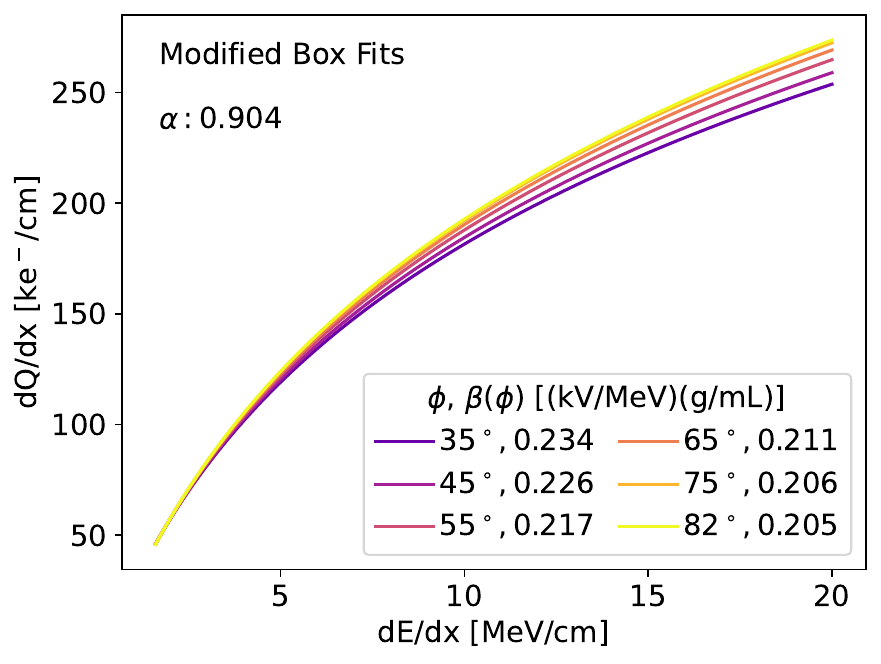}
    \includegraphics[width=0.49\textwidth]{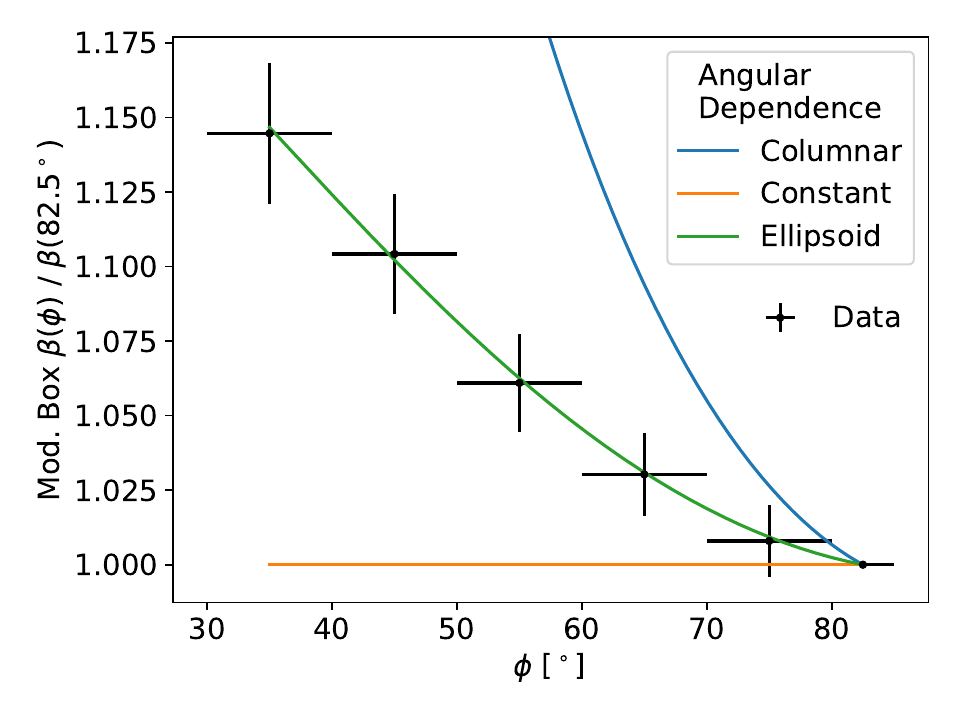}

    \caption{(Left) Modified box fit in each proton angle bin. 
    (Right) Ratio of $\beta(\phi)$ measurements in the modified box fit to the value in the $80^\circ < \phi < 85^\circ$ bin. This ratio is compared to three models of the angular dependence, as described in the text. All three models are normalized to match the data in the $80^\circ < \phi < 85^\circ$ bin.}
    \label{fig:allfit}
\end{figure}

The result of the fit comparing measured $dQ/dx$ and expected $dE/dx$ is shown for muons in figure \ref{fig:muonfits} and for protons in figure \ref{fig:protonfit}. 
The $\chi^2$ values of the modified box and Birks fits are compared for the proton data in table \ref{tbl:fitchi2s} (there is no significant difference for the muon data). The modified box fit results in a lower $\chi^2$ across all proton angular bins. This reduction is greater than can be accounted for by the additional degree of freedom from the extra parameter in the modified box fit ($\alpha$). We use the modified box fit for the energy scale calibration for ICARUS. Figure \ref{fig:allfit} plots all of the modified box fits together, and shows the $\phi$ dependence in the $\beta$ parameter in the fit.

We observe a clear difference in the measured MPV $dQ/dx$ between different proton track angle bins. This can be seen both in the proton data in figure \ref{fig:protonfit}, as well as in the $\beta$ parameter fit in figure \ref{fig:allfit}. At the largest measured $dE/dx$ value (\SI{12}{MeV/cm}), this is a difference of 10\% on the measured proton charge between the largest and smallest $\phi$ bins. We have closely examined any possible angular biases in the charge reconstruction. None were found outside of the 2.5\% correction detailed in section \ref{sec:TrackSelection}. We attribute the angular dependence observed here to an angular dependence in electron-ion recombination in liquid argon. The success of the modified box fits demonstrates that this can be parameterized by including a dependence on the track angle to the drift electric field ($\phi$) in the $\beta$ parameter.

\begin{table}[bh]
\centering
\caption{$\chi^2$ values in the modified box and Birks equation fits to proton data, broken down by the ($\phi$) angle bin. There are 16 data points in each angular bin. The Birks fit has 7 free parameters ($k(\phi), \mathcal{G}$), and the modified box fit has 8 ($\beta(\phi), \alpha, \mathcal{G}$).}

\begin{tabular}{ l | l | l} 
Proton Angle Bin & Birks Fit $\chi^2$ & Modified Box Fit $\chi^2$ \\
\hline
$30^\circ < \phi < 40^\circ$ & $40.6$ & $13.9$\\
$40^\circ < \phi < 50^\circ$ & $57.7$ & $11.7$\\
$50^\circ < \phi < 60^\circ$ & $53.2$ & $6.1$\\
$60^\circ < \phi < 70^\circ$ & $52.3$ & $5.4$\\
$70^\circ < \phi < 80^\circ$ & $52.3$ & $3.9$\\
$80^\circ < \phi < 85^\circ$ & $37.7$ & $4.8$\\
\hline
Total $\chi^2/\text{n d.o.f.}$ & $293.8 / 89$ & $45.8 / 88$\\
\end{tabular}
\label{tbl:fitchi2s}
\end{table}

We investigate the $\phi$ dependence of the $\beta$ parameter by comparing three models of the angular dependence:
\begin{samepage}
\begin{itemize}
    \item Constant: $\beta(\phi) = \beta_{90}$
    \item Columnar \cite{Columnar}: $\beta(\phi) = \beta_{90} / \text{sin}\phi$
    \item Ellipsoid \cite{Ellipsoid}: $\beta(\phi) = \beta_{90} / \sqrt{\text{sin}^2\phi + \text{cos}^2\phi / R^2}$,
\end{itemize}
\end{samepage}
where $\beta_{90}\ (\equiv\beta(90^\circ))$ and $R$ are fit parameters. The $R$ parameter of the ellipsoid model interpolates between the constant ($R = 1$) and columnar ($R \to \infty$) models. Figure \ref{fig:allfit} compares the three models to the measured ratio of $\beta(\phi)$ to $\beta(82.5^\circ)$ (the ratio removes the correlated uncertainty due to the drift electric field). Neither the constant nor columnar models match the dependence. However, the ellipsoid model fit is able to describe it well. 

Putting this together, we find that the ellipsoid modified box (EMB) model of recombination
\begin{equation}
\label{eq:EMB}
    \begin{split}
        \frac{dQ}{dx} &= \frac{\text{log}\left(\alpha + \mathcal{B}(\phi)\frac{dE}{dx}\right)}{\mathcal{B}(\phi)W_\text{ion}}\\
        \mathcal{B}(\phi) &= \frac{\beta_{90}}{\mathcal{E}\rho\sqrt{\text{sin}^2\phi + \text{cos}^2\phi/R^2}}
    \end{split}
\end{equation}
is able to describe the muon and proton data across all measured angles. The ICARUS measurement of the EMB model is obtained by re-fitting the $dQ/dx$ data. The result is

\begin{figure}[t]
    \centering
    \includegraphics[width=0.5\textwidth]{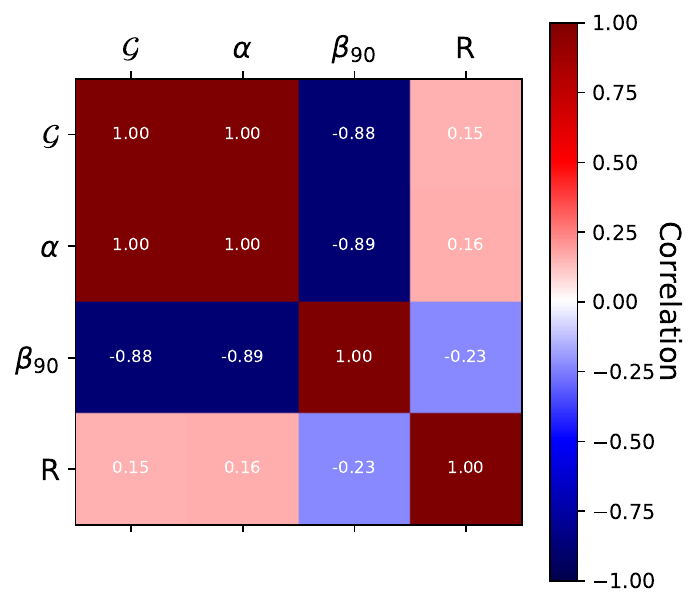}
    \caption{Correlation matrix of the uncertainties in the ellipsoid modified box (EMB) recombination measurement.}
    \label{fig:embcorr}
\end{figure}

\begin{center}    
\begin{tabular}{ll} 
 $\alpha$: 0.904 $\pm$ 0.008 \quad & $R$: 1.25 $\pm$ 0.02\\
\multicolumn{2}{l}{$\beta_{90}$: 0.204 $\pm$ \SI{0.008}{\Bunit},}\\
\end{tabular}
\end{center}
with the ICARUS electronics gain ($\mathcal{G}$) measured as 75.0 $\pm$ 1.1 $e^-$/ADC. Figure \ref{fig:embcorr} displays the correlation matrix of the uncertainties in the fit. The measured $R$ parameter is 12.5 standard deviations away from a value consistent with no angular dependence in recombination ($R=1$).

\begin{figure}[tp]
    \centering
    \includegraphics[width=0.49\textwidth]{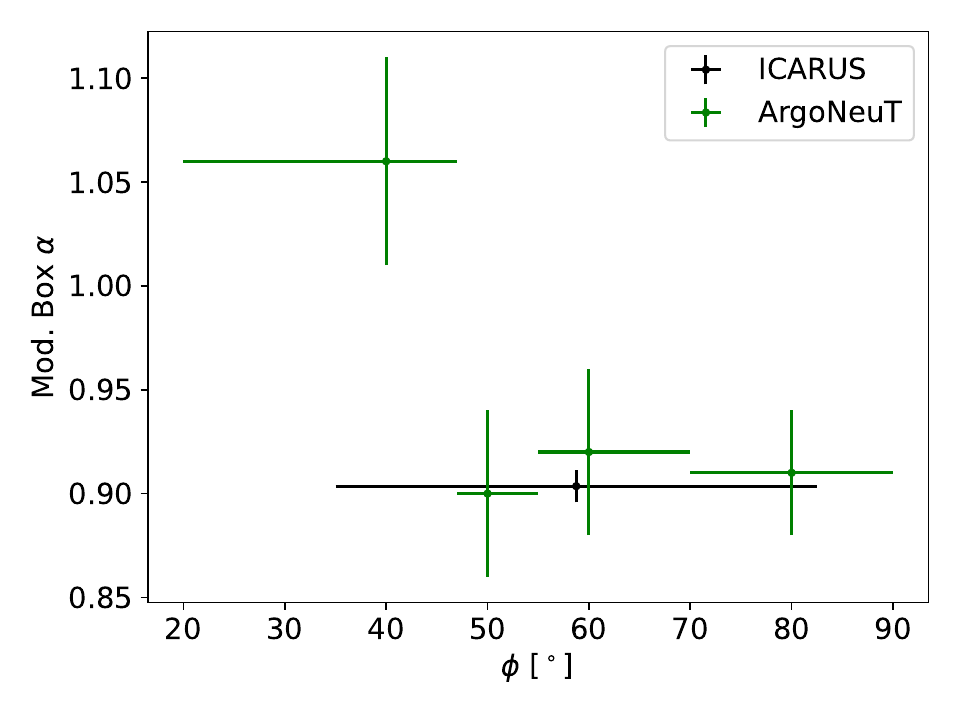}
    \includegraphics[width=0.49\textwidth]{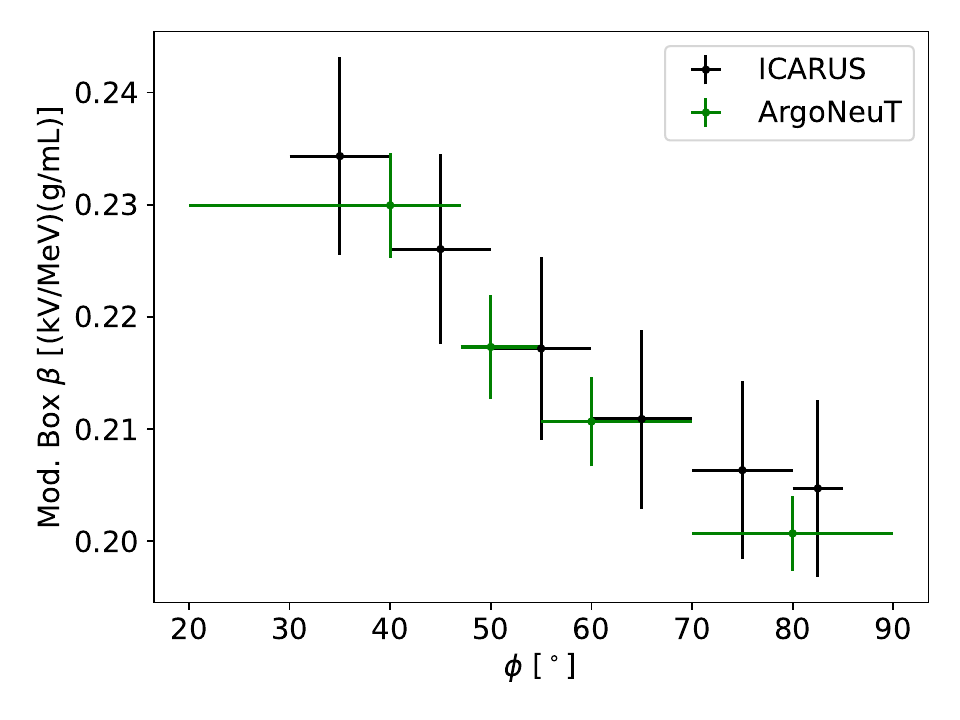}

    \caption{Comparison of the modified box recombination model fit between this measurement and the ArgoNeuT result \cite{NEUTRecomb}. The two fits are not completely comparable because the ArgoNeuT result allowed the $\alpha$ parameter to be different in the different angular bins. Beyond this limitation, the measurements appear consistent.}
    \label{fig:neutcomp}
\end{figure}

The ArgoNeuT experiment previously measured electron-ion recombination, including its angular dependence, in liquid argon with protons and deuterons at a drift field strength close to the field strength in ICARUS \cite{NEUTRecomb}. 
A comparison is shown in figure \ref{fig:neutcomp}. The two measurements appear consistent. While ArgoNeuT recommended its result be applied in an angular independent way, we have found that the angular dependence in the EMB model is critical to properly calibrate the ICARUS LArTPC. Section \ref{sec:ParticleID} demonstrates the impact of the angular dependence on particle identification.


\section{Impact on Particle Identification and Calorimetry}
\label{sec:ParticleID}
\begin{figure}[t]
    \centering
    \includegraphics[width=0.7\textwidth]{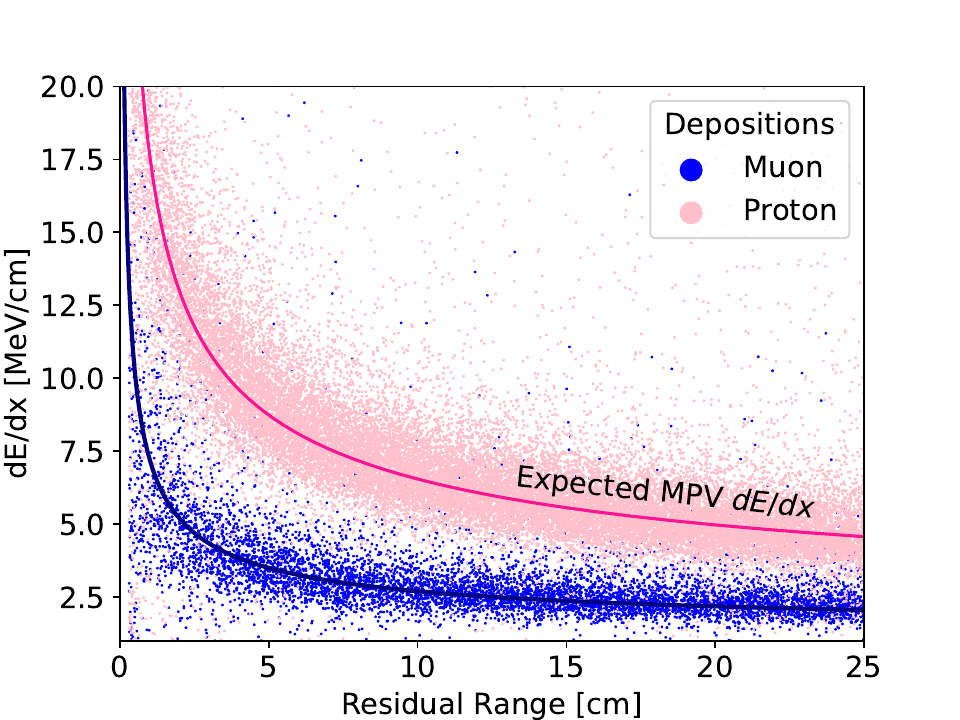}

    \caption{Scatter plot of calibrated energy depositions from selected stopping muons and protons in ICARUS data.}
    \label{fig:mupdep}
\end{figure}

The measurement of ellipsoid modified box (EMB) electron-ion recombination (equation \ref{eq:EMB}) and the ICARUS electronics gain presented here provides the TPC energy scale calibration in ICARUS. This calibration enables the use of calorimetric particle identification (PID) and energy reconstruction for ionizing particle tracks. Figure \ref{fig:mupdep} shows the distribution of calibrated energy depositions for selected stopping muons and protons. The figure demonstrates both the accuracy of the calibration and the ability of the ICARUS TPC to calorimetrically separate muon and proton tracks. 

In this section, the performance of ICARUS calorimetry is compared in detail between data and Monte Carlo simulation. Furthermore, the application in data of the ICARUS EMB-based calibration (equation \ref{eq:EMB}) and an ArgoNeuT modified box-based calibration (equation \ref{eq:modbox}) are compared. The EMB model values are taken from this paper. The ArgoNeuT modified box values are taken from ref.\ \cite{NEUTRecomb}. The electronics gain in the ArgoNeuT modified box-based calibration is determined from a fit to muon $dQ/dx$ data.


\subsection{Particle Identification}

\begin{figure}[t]
    \centering
    \includegraphics[width=\textwidth]{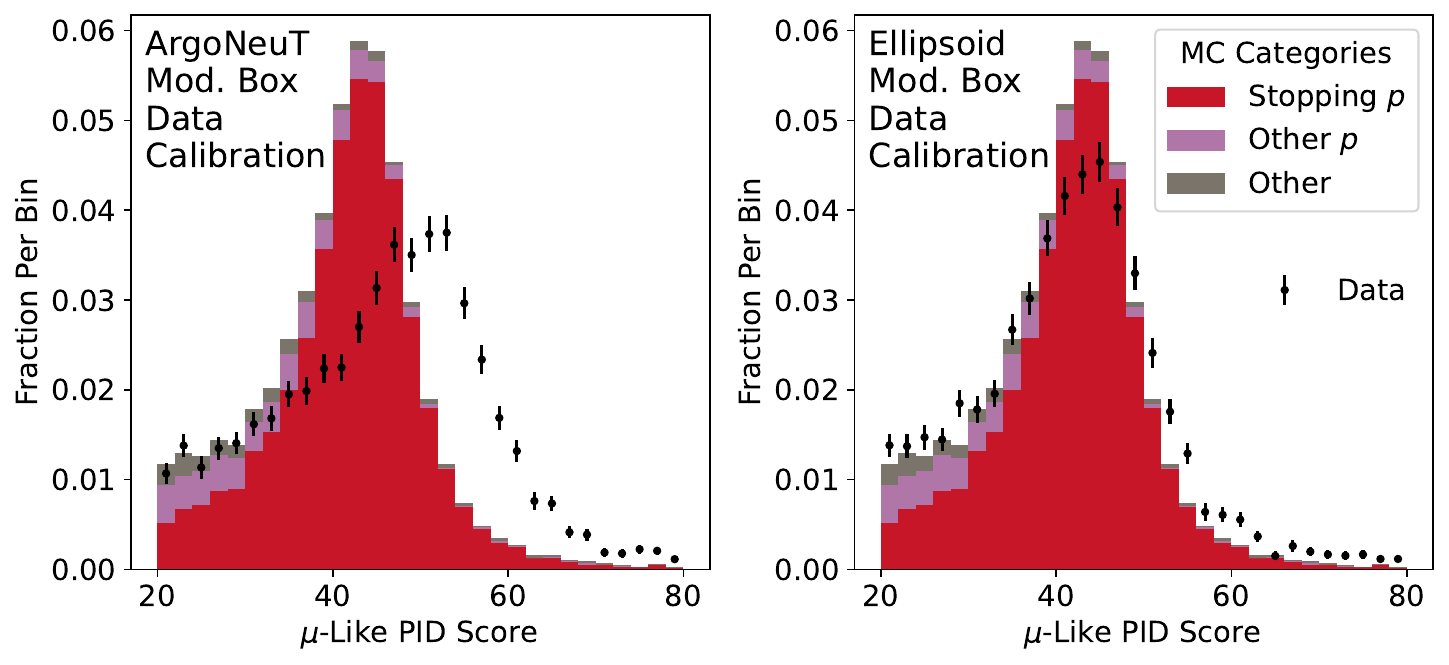}

    \caption{Monte Carlo simulation to data comparison of the \upid score applying the angular independent ArgoNeuT modified box-based calibration (left) and with the angular dependent EMB-based calibration (right). Tracks are selected as detailed in section \ref{sec:ParticleID}. The data is taken with the NuMI beam. The cosmic-triggered component of the data is subtracted with off-beam data.}
    \label{fig:chi2upmcdata}
\end{figure}

Ionization calorimetry can be applied to separate muon and proton tracks with the \upid and \ppid scores, as is shown for ICARUS Monte Carlo simulation in figure \ref{fig:chi2up}. Modeling the distribution of these scores precisely is critical for physics analysis. A data to simulation comparison of the \upid score for proton-like tracks is shown in figure \ref{fig:chi2upmcdata}. Proton-like tracks are selected with the topological track selection detailed in section \ref{sec:protonselection}. They also must have a \ppid score less than 80. The comparison is made for energy depositions in data calibrated with the EMB model (equation \ref{eq:EMB}) and with the angular-independent ArgoNeuT modified box model.

The data in the comparison is taken with the NuMI beam. The cosmic triggered component is removed by subtracting off-beam data, normalized to the trigger livetime. In the Monte Carlo simulation,  NuMI neutrino interactions are modeled by GENIE \cite{GENIE} and cosmic-ray muons are generated by CORSIKA \cite{Corsika}. Generated particles are propagated through the detector by Geant4 \cite{GEANT}. Energy depositions are turned into ionization charge using the angle-independent ArgoNeuT modified box model. Ionization charge is simulated through the ICARUS TPC detector response simulation, which applies the Wire Cell framework \cite{WireCell}, with the simulated field responses tuned to the observed ICARUS TPC signal shapes \cite{ICARUSSignalNoise}. There is a significant uncertainty in the normalization of the Monte Carlo simulation prediction due to neutrino interaction and flux modeling. This uncertainty is mostly removed by the area normalization in the plot. Only statistical uncertainties on the data are shown.

Including the angular dependence in the EMB recombination correction dramatically improves the agreement of the \upid score distributions between data and simulation for stopping protons. There remains some residual disagreement in the broadness of the distribution. A similar effect is also seen in the proton energy reconstruction (see next section).


\subsection{Proton Energy Reconstruction}
\label{sec:protonenergy}

We further examine the performance of calorimetry in ICARUS by comparing the energy reconstructed by range and by calorimetry for stopping protons. This comparison is made for both data and Monte Carlo simulation. As for particle identification, we also compare the impact of applying either the EMB or ArgoNeuT modified box recombination model to calibrate the calorimetric energy reconstruction in data.

Protons are identified using the selection detailed in section \ref{sec:protonselection}. The energy is measured along the last \SI{25}{cm} of the track. The range energy is measured with a lookup table mapping proton length to energy. This table applies the continuous-slowing-down-approximation (CSDA) \cite{StoppingPower}, with the mean energy loss taken from equation \ref{eq:dEdxmean}.

To precisely reconstruct proton calorimetric energy, a so-called ``Q-tip" energy reconstruction technique has been developed. The charge from the last \SI{3}{\centi\meter} of the proton track is summed into a total charge which is converted to a total tip energy (this is based on the method developed by ArgoNeuT for ``blip" energy depositions \cite{ArgoNeuTBlip}). Charge in subsequent hits is corrected for recombination hit-by-hit and then summed into a total energy. This method is more precise than correcting for recombination hit-by-hit along the whole track because charge near the proton tip gets smeared out by diffusion and edge effects. The value of energy loss is also changing rapidly near the proton tip due to its Bragg peak, and so this smearing biases the energy reconstruction at the tip if it is applied hit-by-hit.

\begin{figure}[tp]
    \centering
    \includegraphics[height=0.26\textheight]{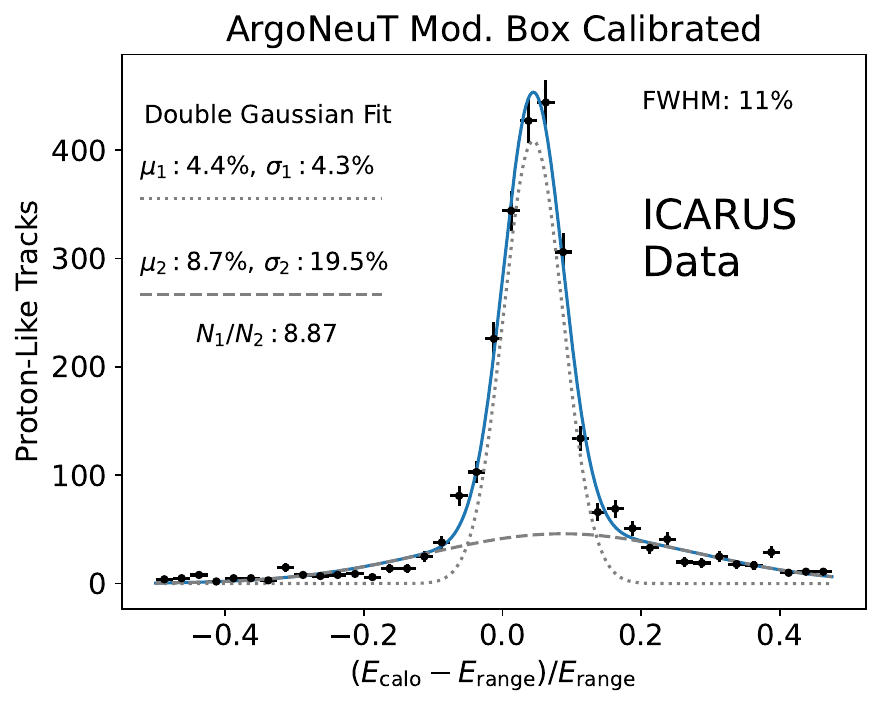}
    \includegraphics[height=0.26\textheight]{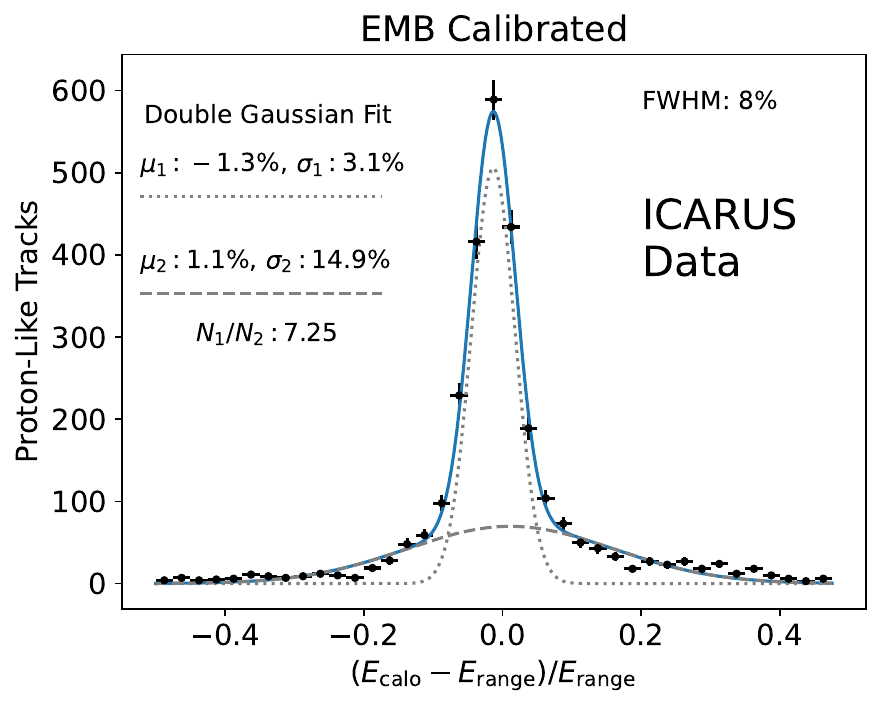}
    \includegraphics[height=0.26\textheight]{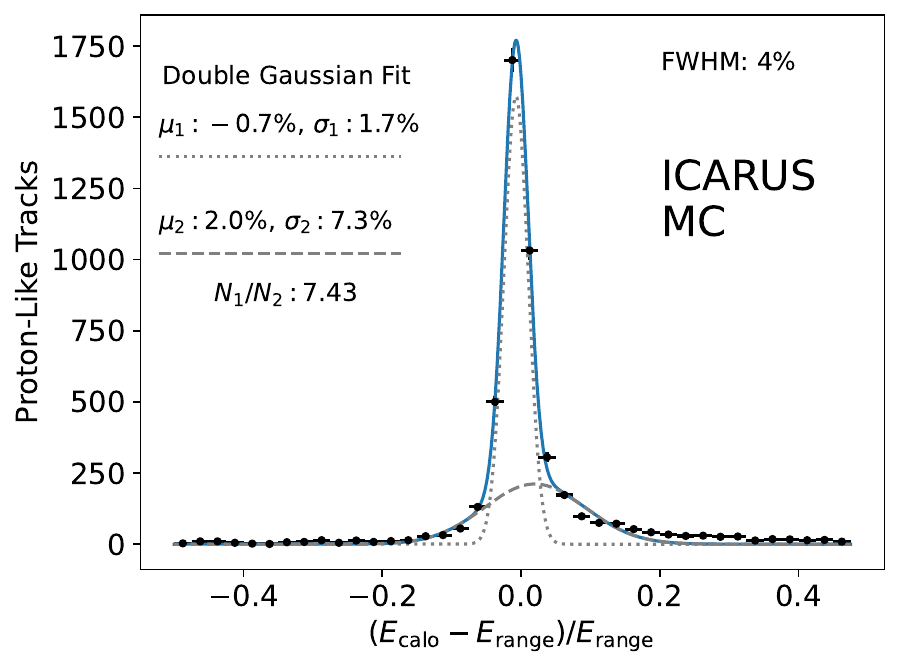}

    \caption{Comparison of calorimetric energy ($E_\text{calo}$) and range energy ($E_\text{range}$) reconstruction (as defined in section \ref{sec:protonenergy}) for selected protons in ICARUS data (top) and Monte Carlo simulation (bottom). The comparison is made in data for the ArgoNeuT modified box-based calibration (top-left) and the EMB-based calibration (top-right). The Monte Carlo simulation applies the ArgoNeuT modified box model for both simulating recombination and correcting for it. The data points are fit to a sum of two Gaussian distributions with centers ($\mu_1$, $\mu_2$) and standard deviations ($\sigma_1$, $\sigma_2$). The ratio of the amplitudes of the Gaussian distributions is quoted as $N_1/N_2$.}
    \label{fig:dataprotonEreco}
\end{figure}

The comparison of the calorimetric and range based energy reconstruction is shown for ICARUS data and Monte Carlo simulation in figure \ref{fig:dataprotonEreco}. Calorimetric energy calibrations applying the ArgoNeuT modified box model and the EMB model are also compared for the data. The result with the EMB-based calibration is both less biased and has a better resolution than for the ArgoNeuT modified box-based calibration.


The full width at half maximum (FWHM) of the relative difference between the range and (EMB calibrated) calorimetric energy values is 8\% in ICARUS data. This resolution is about twice as large in data as in simulation. The resolution of the range-based measurement is likely simulated well by the ICARUS Geant4-based simulation, so the calorimetric energy measurement would be dominating the discrepancy. In principle, this effect could be explained by an underestimation of the inherent charge resolution of the detector in ICARUS simulation. However, as is shown in the next section, for muons (which have mostly minimum ionizing depositions) there is no such disagreement. This points to an effect specific to highly ionizing depositions. In particular, it is possible that fluctuations in ionization recombination are underestimated for highly ionizing depositions in the ICARUS simulation.

\subsection{Muon Energy Reconstruction}
\label{sec:muonenergy}

Stopping muons are identified by the selection from section \ref{sec:muonselection}. The same angular cuts are applied requiring $5^\circ < \theta_{xw} < 20^\circ$ and $70^\circ < \phi < 85^\circ$. The energy is measured along the whole track. The calorimetric energy is computed with the ``Q-tip" energy reconstruction. It also includes a correction for missing hits along the track: for any wire along the track without a reconstructed hit, the mean energy loss at the inferred value of residual range is summed into the calorimetric energy. This correction ameliorates cases where charge depositions along the track go below the signal-to-noise threshold to create hits (especially along the minimum-ionizing part of the muon). It is a 5\% correction on average. 

\begin{figure}[t]
    \centering
    \includegraphics[width=0.485\textwidth]{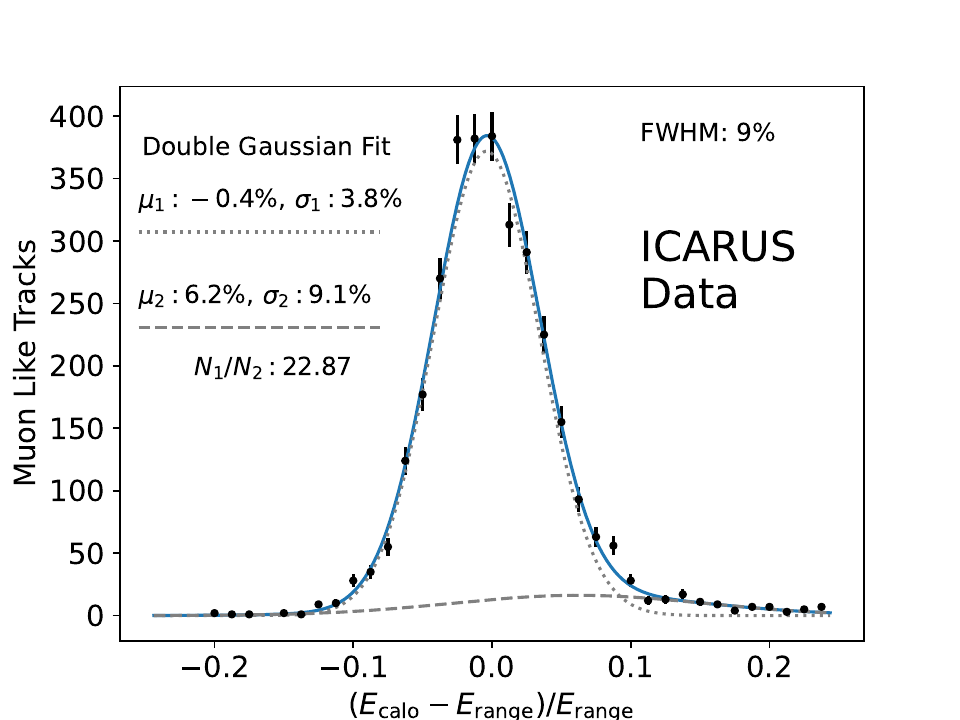}
    \includegraphics[width=0.485\textwidth]{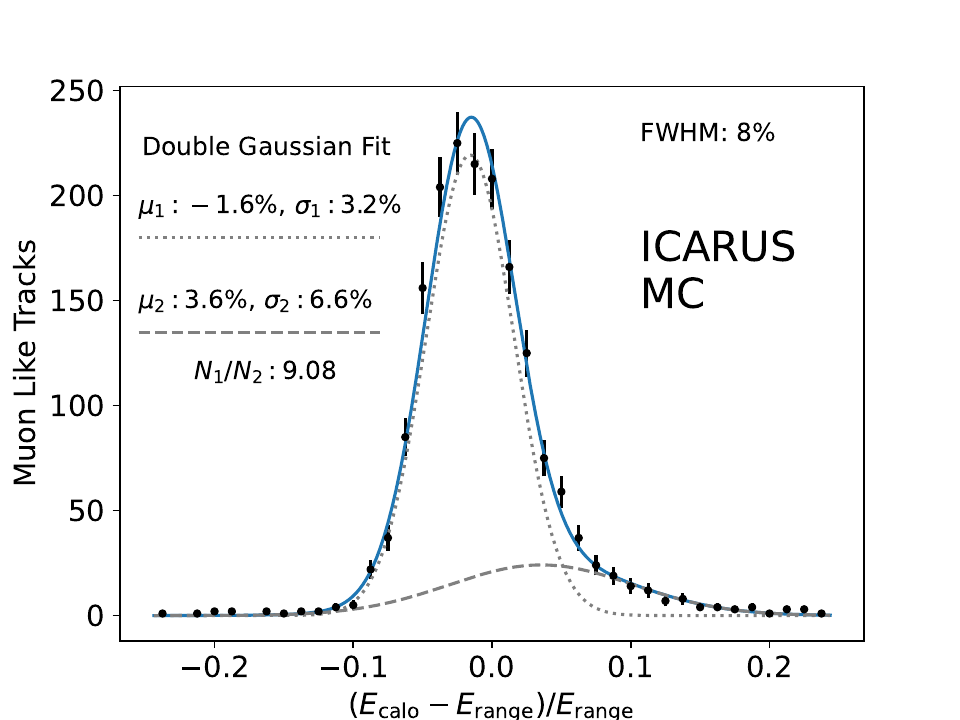}

    \caption{Comparison of calorimetric energy ($E_\text{calo}$) and range energy ($E_\text{range}$) reconstruction for selected muons in ICARUS data (left) and Monte Carlo simulation (right). The EMB-based calibration is applied. The calorimetric energy applies the ``Q-tip" energy reconstruction and a correction for missing hits along the track. The data points are fit to a sum of two Gaussian distributions with centers ($\mu_1$, $\mu_2$) and standard deviations ($\sigma_1$, $\sigma_2$). The ratio of the amplitudes of the Gaussian distributions is quoted as $N_1/N_2$.}
    \label{fig:datamuonEreco}
\end{figure}

Figure \ref{fig:datamuonEreco} shows the comparison of range and calorimetric energy for stopping cosmic-ray muons in ICARUS data and simulation. The comparison to data is only made for the EMB-based calibration because the difference in recombination modelling is not significant for (mostly minimum-ionizing) muon tracks. Unlike in the proton comparison, the distribution of the relative difference in data is described well by ICARUS Monte Carlo simulation.

\section{Conclusion}
\label{sec:Conclusion}
This paper has detailed a new measurement of electron-ion recombination in liquid argon and its application in the energy scale calibration of the ICARUS time projection chamber. This measurement observes a significant angular dependence in recombination for highly-ionizing particles in liquid argon. The ellipsoid modified box (EMB) model  of recombination (equation \ref{eq:EMB}) is able to describe the data across all measured angles.

The recombination measurement is used to calibrate calorimetry in the ICARUS TPC for use in particle identification and energy reconstruction. Distributions of particle identification variables match well between ICARUS data and Monte Carlo simulation when the EMB recombination model is used to calibrate the data. The difference between calorimetric energy reconstruction and the estimation of energy by the track length has a resolution of about 5\% for both muons and protons in ICARUS data. This matches the expectation for muons from ICARUS simulation but is larger than the expectation for protons. This effect could be the result of larger fluctuations in electron-ion recombination for highly ionizing energy depositions than what is simulated.

\section*{Acknowledgements}
\label{sec:acknoweldgements}
This document was prepared by the ICARUS
Collaboration using the resources of the Fermi
National Accelerator Laboratory (Fermilab), a
U.S. Department of Energy, Office of Science,
HEP User Facility. Fermilab is managed by
Fermi Research Alliance, LLC (FRA), acting
under Contract No. DE-AC02-07CH11359.
This work was supported by the US Department of Energy, INFN,EU Horizon 2020 Research
and Innovation Program under the Marie Sklodowska-Curie Grant Agreement
No. 734303, 822185, 858199 and 101003460, and the Horizon Europe Research
and Innovation Program under the Marie Sklodowska-Curie Grant Agreement
No. 101081478
Part of the work resulted from the implementation of
the research Project No. 2019/33/N/ST2/02874
funded by the National Science Centre, Poland.
We also acknowledge the contribution of
many SBND colleagues, in particular for the development of a number of simulation, reconstruction and analysis tools which are shared
within the SBN program.

\bibliographystyle{JHEP}
\bibliography{cite}

\providecommand{\href}[2]{#2}\begingroup\raggedright\begin{thebibliography}{10}

\bibitem{Carlo}
C.~Rubbia, \emph{{The Liquid Argon Time Projection Chamber: A New Concept for Neutrino Detectors}}, {\emph{CERN-EP} {\bfseries 77-08} (1977) }.

\bibitem{ImelRecomb}
J.~Thomas and D.A.~Imel, \emph{Recombination of electron-ion pairs in liquid argon and liquid xenon}, \href{https://doi.org/10.1103/PhysRevA.36.614}{\emph{Phys. Rev. A} {\bfseries 36} (1987) 614}.

\bibitem{ImelRecombStat}
J.~Thomas, D.A.~Imel and S.~Biller, \emph{{Statistics of charge collection in liquid argon and liquid xenon}}, \href{https://doi.org/10.1103/PhysRevA.38.5793}{\emph{Phys. Rev. A} {\bfseries 38} (1988) 5793}.

\bibitem{ICARUSrecomb}
{\scshape ICARUS} collaboration, \emph{{Study of electron recombination in liquid argon with the ICARUS TPC}}, \href{https://doi.org/10.1016/j.nima.2003.11.423}{\emph{Nucl. Instrum. Meth. A} {\bfseries 523} (2004) 275}.

\bibitem{NEUTRecomb}
{\scshape ArgoNeuT} collaboration, \emph{{A Study of Electron Recombination Using Highly Ionizing Particles in the ArgoNeuT Liquid Argon TPC}}, \href{https://doi.org/10.1088/1748-0221/8/08/P08005}{\emph{JINST} {\bfseries 8} (2013) P08005} [\href{https://arxiv.org/abs/1306.1712}{{\ttfamily 1306.1712}}].

\bibitem{Columnar}
G.~Jaffé, \emph{Zur theorie der ionisation in kolonnen}, \href{https://doi.org/10.1002/andp.19133471205}{\emph{Annalen der Physik} {\bfseries 347} (1913) 303 }.

\bibitem{Ellipsoid}
V.~Cataudella, A.~de~Candia, G.D.~Filippis, S.~Catalanotti, M.~Cadeddu, M.~Lissia et~al., \emph{{Directional modulation of electron-ion pairs recombination in liquid argon}}, \href{https://doi.org/10.1088/1748-0221/12/12/P12002}{\emph{JINST} {\bfseries 12} (2017) P12002}.

\bibitem{ArDM}
M.~Cadeddu et~al., \emph{{Directional dark matter detection sensitivity of a two-phase liquid argon detector}}, \href{https://doi.org/10.1088/1475-7516/2019/01/014}{\emph{JCAP} {\bfseries 01} (2019) 014} [\href{https://arxiv.org/abs/1704.03741}{{\ttfamily 1704.03741}}].

\bibitem{ICARUSOG}
{\scshape ICARUS} collaboration, \emph{{Design, construction and tests of the ICARUS T600 detector}}, \href{https://doi.org/10.1016/j.nima.2004.02.044}{\emph{Nucl. Instrum. Meth. A} {\bfseries 527} (2004) 329}.

\bibitem{ICARUSOverhaul}
{\scshape ICARUS-T600} collaboration, \emph{{Overhaul and Installation of the ICARUS-T600 Liquid Argon TPC Electronics for the FNAL Short Baseline Neutrino Program}}, \href{https://doi.org/10.1088/1748-0221/16/01/P01037}{\emph{JINST} {\bfseries 16} (2021) P01037} [\href{https://arxiv.org/abs/2010.02042}{{\ttfamily 2010.02042}}].

\bibitem{NuMI}
P.~Adamson et~al., \emph{{The NuMI Neutrino Beam}}, \href{https://doi.org/10.1016/j.nima.2015.08.063}{\emph{Nucl. Instrum. Meth. A} {\bfseries 806} (2016) 279} [\href{https://arxiv.org/abs/1507.06690}{{\ttfamily 1507.06690}}].

\bibitem{SBNProposal}
{\scshape MicroBooNE, LAr1-ND, ICARUS-WA104} collaboration, \emph{{A Proposal for a Three Detector Short-Baseline Neutrino Oscillation Program in the Fermilab Booster Neutrino Beam}},  \href{https://arxiv.org/abs/1503.01520}{{\ttfamily 1503.01520}}.

\bibitem{SBNProgram}
P.A.~Machado, O.~Palamara and D.W.~Schmitz, \emph{{The Short-Baseline Neutrino Program at Fermilab}}, \href{https://doi.org/10.1146/annurev-nucl-101917-020949}{\emph{Ann. Rev. Nucl. Part. Sci.} {\bfseries 69} (2019) 363} [\href{https://arxiv.org/abs/1903.04608}{{\ttfamily 1903.04608}}].

\bibitem{ICARUSElectronics}
{\scshape ICARUS/NP01} collaboration, \emph{{New read-out electronics for ICARUS-T600 liquid Argon TPC. Description, simulation and tests of the new front-end and ADC system}}, \href{https://doi.org/10.1088/1748-0221/13/12/P12007}{\emph{JINST} {\bfseries 13} (2018) P12007} [\href{https://arxiv.org/abs/1805.03931}{{\ttfamily 1805.03931}}].

\bibitem{ICARUSResults}
{\scshape ICARUS} collaboration, \emph{{ICARUS at the Fermilab Short-Baseline Neutrino program: initial operation}}, \href{https://doi.org/10.1140/epjc/s10052-023-11610-y}{\emph{Eur. Phys. J. C} {\bfseries 83} (2023) 467} [\href{https://arxiv.org/abs/2301.08634}{{\ttfamily 2301.08634}}].

\bibitem{ICARUSSignalNoise}
{\scshape ICARUS} collaboration, \emph{{Calibration and simulation of ionization signal and electronics noise in the ICARUS liquid argon time projection chamber}},  \href{https://arxiv.org/abs/2407.11925}{{\ttfamily 2407.11925}}.

\bibitem{Bethe}
H.~{Bethe}, \emph{{Zur Theorie des Durchgangs schneller Korpuskularstrahlen durch Materie}}, \href{https://doi.org/10.1002/andp.19303970303}{\emph{Annalen der Physik} {\bfseries 397} (1930) 325}.

\bibitem{PDG}
P.A.Z.~et.~al. (Particle Data~Group), \emph{{Review of Particle Physics}}, \href{https://doi.org/10.1093/ptep/ptaa104}{\emph{Progress of Theoretical and Experimental Physics} {\bfseries 2020} (2020) }.

\bibitem{DensityEffect}
R.~Sternheimer, M.~Berger and S.~Seltzer, \emph{Density effect for the ionization loss of charged particles in various substances}, \href{https://doi.org/https://doi.org/10.1016/0092-640X(84)90002-0}{\emph{Atomic Data and Nuclear Data Tables} {\bfseries 30} (1984) 261}.

\bibitem{StoppingPower}
D.E.~Groom, N.V.~Mokhov and S.I.~Striganov, \emph{Muon stopping power and range tables 10 mev–100 tev}, \href{https://doi.org/https://doi.org/10.1006/adnd.2001.0861}{\emph{Atomic Data and Nuclear Data Tables} {\bfseries 78} (2001) 183}.

\bibitem{Landau}
L.~Landau, \emph{{On the energy loss of fast particles by ionization}}, {\emph{J. Phys. (USSR)} {\bfseries 8} (1944) 201}.

\bibitem{Vavilov}
P.V.~Vavilov, \emph{{Ionization losses of high-energy heavy particles}}, {\emph{Sov. Phys. JETP} {\bfseries 5} (1957) 749}.

\bibitem{GrayDiff}
G.~Putnam and D.W.~Schmitz, \emph{{Effect of diffusion on the peak value of energy loss observed in a LArTPC}}, \href{https://doi.org/10.1088/1748-0221/17/10/P10044}{\emph{JINST} {\bfseries 17} (2022) P10044} [\href{https://arxiv.org/abs/2205.06745}{{\ttfamily 2205.06745}}].

\bibitem{MichelleDiff}
A.~Lister and M.~Stancari, \emph{{Investigations on a fuzzy process: effect of diffusion on calibration and particle identification in Liquid Argon Time Projection Chambers}}, \href{https://doi.org/10.1088/1748-0221/17/07/P07016}{\emph{JINST} {\bfseries 17} (2022) P07016} [\href{https://arxiv.org/abs/2201.09773}{{\ttfamily 2201.09773}}].

\bibitem{ROOT}
R.~Brun and F.~Rademakers, \emph{Root - an object oriented data analysis framework},  in \emph{AIHENP'96 Workshop, Lausane}, vol.~389, pp.~81--86, 1996.

\bibitem{Strait}
M.~Strait, \emph{Evaluation of the mean excitation energies of gaseous and liquid argon}, \href{https://doi.org/10.1088/1748-0221/19/01/P01009}{\emph{Journal of Instrumentation} {\bfseries 19} (2024) P01009}.

\bibitem{Wion}
M.~Miyajima, T.~Takahashi, S.~Konno, T.~Hamada, S.~Kubota, H.~Shibamura et~al., \emph{{Average energy expended per ion pair in liquid argon}}, \href{https://doi.org/10.1103/PhysRevA.9.1438}{\emph{Phys. Rev. A} {\bfseries 9} (1974) 1438}.

\bibitem{RecombinationGerminate}
M.~Wojcik and M.~Tachiya, \emph{Electron thermalization and electron–ion recombination in liquid argon}, \href{https://doi.org/https://doi.org/10.1016/j.cplett.2003.08.006}{\emph{Chemical Physics Letters} {\bfseries 379} (2003) 20}.

\bibitem{RecombinationThermal}
U.~Sowada, J.M.~Warman and M.P.~de~Haas, \emph{Hot-electron thermalization in solid and liquid argon, krypton, and xenon}, \href{https://doi.org/10.1103/PhysRevB.25.3434}{\emph{Phys. Rev. B} {\bfseries 25} (1982) 3434}.

\bibitem{Pandora}
{\scshape MicroBooNE} collaboration, \emph{{The Pandora multi-algorithm approach to automated pattern recognition of cosmic-ray muon and neutrino events in the MicroBooNE detector}}, \href{https://doi.org/10.1140/epjc/s10052-017-5481-6}{\emph{Eur. Phys. J. C} {\bfseries 78} (2018) 82} [\href{https://arxiv.org/abs/1708.03135}{{\ttfamily 1708.03135}}].

\bibitem{DUNEPandora}
{\scshape DUNE} collaboration, \emph{{Reconstruction of interactions in the ProtoDUNE-SP detector with Pandora}}, \href{https://doi.org/10.1140/epjc/s10052-023-11733-2}{\emph{Eur. Phys. J. C} {\bfseries 83} (2023) 618} [\href{https://arxiv.org/abs/2206.14521}{{\ttfamily 2206.14521}}].

\bibitem{Corsika}
D.~Heck, J.~Knapp, J.N.~Capdevielle, G.~Schatz and T.~Thouw, \emph{{CORSIKA: A Monte Carlo code to simulate extensive air showers}}, {\emph{FZKA} {\bfseries 6019} (1998) }.

\bibitem{ArgoNeuTPID}
C.~Anderson et~al., \emph{{The ArgoNeuT Detector in the NuMI Low-Energy beam line at Fermilab}}, \href{https://doi.org/10.1088/1748-0221/7/10/P10019}{\emph{JINST} {\bfseries 7} (2012) P10019} [\href{https://arxiv.org/abs/1205.6747}{{\ttfamily 1205.6747}}].

\bibitem{CalibrationuB}
{\scshape MicroBooNE} collaboration, \emph{{Calibration of the charge and energy loss per unit length of the MicroBooNE liquid argon time projection chamber using muons and protons}}, \href{https://doi.org/10.1088/1748-0221/15/03/P03022}{\emph{JINST} {\bfseries 15} (2020) P03022} [\href{https://arxiv.org/abs/1907.11736}{{\ttfamily 1907.11736}}].

\bibitem{GENIE}
{\scshape GENIE} collaboration, \emph{{Recent highlights from GENIE v3}}, \href{https://doi.org/10.1140/epjs/s11734-021-00295-7}{\emph{Eur. Phys. J. ST} {\bfseries 230} (2021) 4449} [\href{https://arxiv.org/abs/2106.09381}{{\ttfamily 2106.09381}}].

\bibitem{GEANT}
S.~Agostinelli et~al., \emph{Geant4—a simulation toolkit}, \href{https://doi.org/https://doi.org/10.1016/S0168-9002(03)01368-8}{\emph{Nuclear Instruments and Methods in Physics Research Section A: Accelerators, Spectrometers, Detectors and Associated Equipment} {\bfseries 506} (2003) 250}.

\bibitem{WireCell}
{\scshape MicroBooNE} collaboration, \emph{{Ionization electron signal processing in single phase LArTPCs. Part I. Algorithm Description and quantitative evaluation with MicroBooNE simulation}}, \href{https://doi.org/10.1088/1748-0221/13/07/P07006}{\emph{JINST} {\bfseries 13} (2018) P07006} [\href{https://arxiv.org/abs/1802.08709}{{\ttfamily 1802.08709}}].

\bibitem{ArgoNeuTBlip}
{\scshape ArgoNeuT} collaboration, \emph{{Demonstration of MeV-Scale Physics in Liquid Argon Time Projection Chambers Using ArgoNeuT}}, \href{https://doi.org/10.1103/PhysRevD.99.012002}{\emph{Phys. Rev. D} {\bfseries 99} (2019) 012002} [\href{https://arxiv.org/abs/1810.06502}{{\ttfamily 1810.06502}}].

\end{thebibliography}\endgroup







\end{document}